\documentclass[10pt,letterpaper]{article}
\usepackage[T1]{fontenc}
\usepackage[latin9]{inputenc}
\usepackage[letterpaper]{geometry}
\usepackage{amsmath}
\usepackage{graphicx}
\usepackage{setspace}
\usepackage{amssymb}
\usepackage{esint}
\makeatletter

\providecommand{\tabularnewline}{\\}

\pdfoutput=1

\usepackage[T1]{fontenc}
\usepackage{ae}
\usepackage{aecompl}

\usepackage[english]{babel}

\usepackage[sf,bf,small]{titlesec}

\usepackage[small,sf,bf,hang]{caption}

\long\def\symbolfootnote[#1]#2{\begingroup%
\def\thefootnote{\fnsymbol{footnote}}\footnote[#1]{#2}\endgroup}

\newcommand{\newstar}{\includegraphics[height=1.5em]{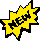}\hspace{-0.5em} }

\newcommand{\tablespacer}{\vspace{1mm}}

\usepackage[nosort]{cite} 

\usepackage{collect}

\renewcommand{\vec}[1]{{\bf #1}}

\newcommand{\nosum}{\Sigma \mspace{-10mu} \big/ }

\DeclareMathOperator{\STr}{STr}

\DeclareMathOperator{\diag}{diag} 

\DeclareMathOperator{\re}{Re}  
\DeclareMathOperator{\im}{Im}

\DeclareMathOperator{\sech}{sech}

\AtBeginDocument{
  
}

\makeatother

\begin{document}
\begin{flushright}
arXiv:0903.3365\\
BROWN-HET 1580\\
UUITP-11/09\bigskip\bigskip
\par\end{flushright}

\begin{center}
\textsf{\textbf{\Large Dyonic Giant Magnons in $CP^{3}$: Strings
and Curves at Finite $J$}}
\par\end{center}{\Large \par}

\begin{center}
\bigskip Michael C. Abbott,$^{b}$ In\^{e}s Aniceto$^{b}$ and Olof
Ohlsson Sax$^{u}$\emph{ \smallskip }\\
\emph{$^{b}$ Brown University, Providence, RI, USA}\\
\emph{$^{u}$ Uppsala University, Uppsala, Sweden}\\
\emph{\smallskip abbott, nes@het.brown.edu \& olof.ohlsson-sax@physics.uu.se}
\par\end{center}

\begin{center}
\bigskip  19 March 2009
\par\end{center}
\begin{quote}
\bigskip \bigskip \textsf{\textbf{Abstract:}} This paper studies
giant magnons in $AdS_{4}\times CP^{3}$ using both the string sigma-model
and the algebraic curve. We complete the dictionary of solutions by
finding the dyonic generalisation of the $CP^{1}$ string solution,
which matches the `small' giant magnon in the algebraic curve, and
by pointing out that the solution recently constructed by the dressing
method is the `big' giant magnon. We then use the curve to compute
finite-$J$ corrections to all cases, which for the non-dyonic cases
always match the AFZ result. For the dyonic $RP^{3}$ magnon we recover
the $S^{5}$ answer, but for the `small' and `big' giant magnons we
obtain new corrections. \bigskip  \thispagestyle{empty} 
\end{quote}

\section{Introduction}

Classical and semiclassical strings allow us to explore some sectors
of the $\mathcal{N}=6$ ABJM / $AdS_{4}\times CP^{3}$ duality \cite{Aharony:2008ug},
and these are much richer than their well-known counterparts in the
$\mathcal{N}=4$ SYM / $AdS_{5}\times S^{5}$ duality \cite{Maldacena:1997re}.
It was known very early on that there are at least two kinds of giant
magnons, created by placing the HM giant magnon \cite{Hofman:2006xt}
into various $S^{2}$-like subspaces, namely $CP^{1}$ and $RP^{2}$
\cite{Gaiotto:2008cg,Grignani:2008is}. It is equally easy to place
Dorey's $S^{3}$ dyonic giant magnon \cite{Dorey:2006dq,Chen:2006gea}
into $RP^{3}$, giving a two-spin generalisation of the $RP^{2}$
magnon \cite{Ahn:2008hj}.

Solutions of the string sigma-model should be in exact correspondence
to algebraic curves \cite{Gromov:2008bz}. Here too several giant
magnon solutions were known (compared to one in the $S^{5}$ case
\cite{Minahan:2006bd}) named `small' and `big' \cite{Shenderovich:2008bs}.
However these could not be the same two solutions as those known in
the sigma-model. In the $S^{5}$ case, and for the small magnon, one
naturally obtains a dyonic (two-parameter, two-spin) solution. But
the big magnon is something not seen in the $S^{5}$ case, a two-parameter
solution with only one non-zero angular momentum, and thus cannot
be the $RP^{3}$ magnon. There are also two distinct small giant magnons,
and it has been observed that a pair of small magnons has all the
properties we expect the $RP^{3}$ magnon to have. \cite{Lukowski:2008eq}

The situation has improved with the recent publication of a new string
solution, found using the dressing method, which like the big giant
magnon is a two-parameter one-angular-momentum solution \cite{Hollowood:2009tw,Kalousios:2009mp,Suzuki:2009sc}.
They have exactly the same dispersion relation, and in both cases
we can take a non-dyonic limit and recover the $RP^{2}$ / pair of
small magnons.

In this paper we complete the puzzle by finding a dyonic generalisation
of the $CP^{1}$ magnon. This is a solution which does not exist in
$S^{5}$, exploring the four-dimensional subspace $CP^{2}$, and it
has a dispersion relation matching that of the small giant magnon.
It exists in two orientations, and like the small magnon has a third
angular momentum which is $J_{3}=\pm Q$ in these two cases. We have
as yet only been able to find the $p=\pi$ case of this solution,
but this is sufficient to see these properties.

Finite-$J$ corrections are of increasing importance in the study
of gauge and string integrability. They can sometimes be computed
directly on the string side by finding solutions with $J<\infty$,
and all existing finite-$J$ giant magnons are embeddings of well-known
$S^{5}$ results of this type \cite{Arutyunov:2006gs,Okamura:2006zv,Astolfi:2007uz,Hatsuda:2008gd}.
Other methods that have been used to calculate finite size corrections
includes the construction of corresponding algebraic curves \cite{Minahan:2008re,Gromov:2008ec,Sax:2008in,Lukowski:2008eq}
and the Lüscher formulae \cite{Janik:2007wt,Lukowski:2008eq,Hatsuda:2008gd,Bombardelli:2008qd}.

In this paper we extend the algebraic curve calculations of \cite{Lukowski:2008eq},
by calculating finite-$J$ corrections not only for a pair of small
giant magnons, but also for a single small magnon and the big magnon.
In the non-dyonic case, all of these give the result AFZ \cite{Arutyunov:2006gs}
found for a magnon in $S^{2}$. Likewise the dyonic pair of giant
magnons matches the $S^{3}$ result: this too is a simple embedding
of that string solution. But for the dyonic small and `dyonic' big
magnons, which correspond to string solutions not found in $S^{5}$,
we find new formulae for these energy corrections.%
\footnote{A word about terminology. We use dyonic to mean two-parameter two-charge
solutions (like Dorey's) but sometimes write `dyonic' (with scare
quotes) for the two-parameter one-charge solution to specify that
we mean the case $r\neq1$. When we speak of the non-dyonic limit,
we always mean that we take the second parameter $r\to1$, and this
always takes us to some embedding of the simplest HM magnon.%
}

\subsection*{Outline}

In section \ref{sec:The-string-sigma-model} we set up the string
sigma-model, and discuss embeddings of $S^{5}$ giant magnons, including
their finite-$J$ corrections. We also discuss the recently published
dressing method solution. Then in section \ref{sec:CP1-Dyonic} we
construct the dyonic generalisation of the $CP^{1}$ magnon, which
lives in $CP^{2}$. 

We then turn to the algebraic curve, and in section \ref{sec:The-algebraic-curve}
set this up, and discuss the known giant magnons in this formalism.
In section \ref{sec:Curve-Finite-J} we calculate finite-$J$ corrections
to all of these solutions. These corrections match the string results
wherever they are known, and are new results in the dyonic small and
big cases. 

Most of our results are summarised in two tables, on pages \pageref{tab:String-Table}
and \pageref{tab:Curve-Table}.

\section{The string sigma-model for $AdS_{4}\times CP^{3}$\label{sec:The-string-sigma-model}}

The string dual of ABJM theory in the 't Hooft limit is type IIA superstrings
in $AdS_{4}\times CP^{3}$. At strong coupling in the gauge theory,
leading to classical strings, these have large radii $R/2$ and $R$.
The metric is then \[
ds^{2}=\frac{R^{2}}{4}ds_{AdS}^{2}+R^{2}ds_{CP}^{2}=R^{2}\left(\frac{dy_{\mu}dy^{\mu}}{-4\vec{y}^{2}}+\frac{dz_{i}d\bar{z}_{i}}{\left|\vec{z}\right|^{2}}-\frac{\left|z_{i}d\bar{z}_{i}\right|^{2}}{\left|\vec{z}\right|^{4}}\right)\]
where we have embedded $AdS_{4}\subset\mathbb{R}^{2,4}$ and $CP^{3}\subset\mathbb{C}^{4}$
, parameterised by $\vec{y}$ and $\vec{z}$ respectively. To study
strings in this space, we constrain the lengths of these embedding
co-ordinate vectors: $\vec{y}^{2}=y_{\mu}y^{\mu}=-(y^{-1})^{2}-(y^{0})^{2}+(y^{1})^{2}+(y^{2})^{2}+(y^{3})^{2}=-1$
and $\left|\vec{z}\right|^{2}=z_{1}\bar{z}_{1}+z_{2}\bar{z}_{2}+z_{3}\bar{z}_{3}=+1$.
In addition to these constraints, points in $\mathbb{C}^{4}$ differing
by an overall phase are identified in $CP^{3}$. This can be dealt
with by introducing a gauge field: write the conformal gauge Lagrangian
as\[
2\mathcal{L}=\frac{1}{4}\partial_{a}\vec{y}\cdot\partial^{a}\vec{y}-\Lambda(\vec{y}^{2}+1)+\overline{D_{a}\vec{z}}\cdot D^{a}\vec{z}-\Lambda'(\bar{\vec{z}}\cdot\vec{z}-1)\]
where the covariant derivative is $D_{a}=\partial_{a}-A_{a}$. The
equation of motion for the gauge field fixes $A_{a}=\bar{\vec{z}}\cdot\partial_{a}\vec{z}$.
We can write the equations of motion for $\vec{y}$ and $\vec{z}$
as%
\footnote{Note that the $CP^{3}$ equation here reduces to that derived in \cite{Abbott:2008qd},
where instead of treating the total phase as a gauge symmetry it was
fixed to a constant using another Lagrange multiplier.%
}\begin{align*}
\partial_{a}\partial^{a}\vec{y}+(\partial_{a}\vec{y}\cdot\partial^{a}\vec{y})\vec{y} & =0, & D_{a}D^{a}\vec{z}+\left(\overline{D_{a}\vec{z}}\cdot D^{a}\vec{z}\right)\vec{z} & =0.\end{align*}
The $AdS$ and $CP$ components are coupled by the Virasoro constraints,
which read:\begin{align*}
\frac{1}{4}\partial_{t}\vec{y}\cdot\partial_{t}\vec{y}+\overline{D_{t}\vec{z}}\cdot D_{t}\vec{z}+\frac{1}{4}\partial_{x}\vec{y}\cdot\partial_{x}\vec{y}+\overline{D_{x}\vec{z}}\cdot D_{x}\vec{z} & =0\\
\frac{1}{4}\partial_{t}\vec{y}\cdot\partial_{x}\vec{y}+\re\left(\overline{D_{t}\vec{z}}\cdot D_{x}\vec{z}\right) & =0.\end{align*}

We now restrict to solutions in $\mathbb{R}\times CP^{3}$, with $y^{-1}+iy^{0}=e^{2it}$
and $y^{1}=y^{2}=y^{3}=0$. We will always work in a gauge in which
this $t$ is worldsheet time (timelike, or static, conformal gauge).%
\footnote{This implies the the length of the worldsheet cannot be held fixed
to $2\pi$. Instead it is proportional to the energy $\Delta$, and
thus infinite for the giant magnon. Taking this to be finite makes
$\Delta$ and $J$ finite too, thus we use `finite-$J$' and `finite-size'
interchangeably. %
} The metric then reduces to \[
ds_{\mathbb{R}\times CP^{3}}^{2}=-dt^{2}+\left|d\vec{z}\right|^{2}-\left|\bar{\vec{z}}\cdot d\vec{z}\right|^{2}.\]
In writing the Lagrangian, and this metric, we have pulled out the
large radius factor $R^{2}=2^{5/2}\pi\sqrt{\lambda}$ to give a prefactor
to the action:

\[
S=\int\frac{dx\, dt}{2\pi}\, R^{2}\mathcal{L}=2\sqrt{2\lambda}\int dx\, dt\,\mathcal{L}.\]
This same factor appears when calculating conserved charges. The one
from time-translation (which we define with respect to $AdS$ time,
$\tan\tau_{AdS}=y^{0}/y^{-1}$) is simply \[
\Delta=2\sqrt{2\lambda}\int dx\:\frac{\partial\mathcal{L}}{\partial\,\partial_{t}\tau_{AdS}}=\sqrt{2\lambda}\int dx\:1\]
where at the end we use the fact that $\tau_{AdS}=2t$ for the the
solutions we're studying. The charges from rotations of $CP^{3}$'s
embedding co-ordinate planes are: \begin{align*}
J(z_{i}) & =2\sqrt{2\lambda}\int dx\:\frac{\partial\mathcal{L}}{\partial\,\partial_{t}(\arg Z_{i})}=2\sqrt{2\lambda}\int dx\left[\im(\bar{z}_{i}\partial_{t}z_{i})-\left|z_{i}\right|^{2}\sum_{j}\im(\bar{z}_{j}\partial_{t}z_{j})\right]\\
 & =2\sqrt{2\lambda}\int dx\im\left(\bar{z}_{i}D_{t}z_{i}\right)\qquad(\nosum_{i})\end{align*}
Only three of the four $J(z_{i})$ are independent, since $\sum_{i=1}^{4}J(z_{i})=0$.
These three are the charges from the Cartan generators of $su(4)$,
and the charges from all of the generators can be obtained using their
Lie-algebra matrices $T^{a}=(T^{a})_{ij}$:\[
J[T^{a}]=2\sqrt{2\lambda}\int dx\im\left(\bar{\vec{z}}\cdot T^{a}D_{t}\vec{z}\right)\]
The matrices $T^{a}$ are Hermitian and traceless, and the charges
$J(z_{i})$ are those generated by diagonal $T^{a}$. The charges
we will need for the giant magnons are\begin{align}
J & =J(z_{1})-J(z_{4})=J\left[\diag(1,0,0,-1)\right]\nonumber \\
Q & =J(z_{2})-J(z_{3})=J\left[\diag(0,1,-1,0)\right]\label{eq:eq:defn-JandQ}\\
J_{3} & =J\left[\diag(-\tfrac{1}{2},\tfrac{1}{2},\tfrac{1}{2},-\tfrac{1}{2})\right]\nonumber \end{align}

Instead of using four complex numbers as co-ordinates for $CP^{3}$,
we can use six real angles. One set of these is defined by \begin{align}
\vec{z} & =\left(\begin{array}{l}
\sin\xi\:\cos(\vartheta_{2}/2)\: e^{-i\eta/2}\: e^{i\varphi_{2}/2}\\
\cos\xi\:\cos(\vartheta_{1}/2)\: e^{i\eta/2}\: e^{i\varphi_{1}/2}\\
\cos\xi\:\sin(\vartheta_{1}/2)\: e^{i\eta/2}\: e^{-i\varphi_{1}/2}\\
\sin\xi\:\sin(\vartheta_{2}/2)\: e^{-i\eta/2}\: e^{-i\varphi_{2}/2}\end{array}\right)\label{eq:defn-6-angles}\end{align}
and in terms of these angles, the metric is:\begin{align}
ds_{CP^{3}}^{2} & =d\xi^{2}+\frac{1}{4}\sin^{2}2\xi\left(d\eta+\frac{1}{2}\cos\vartheta_{1}\: d\varphi_{1}-\frac{1}{2}\cos\vartheta_{2}\: d\varphi_{2}\right)^{2}\nonumber \\
 & \qquad+\frac{1}{4}\cos^{2}\xi\left(d\vartheta_{1}^{2}+\sin^{2}\vartheta_{1}\: d\varphi_{1}^{2}\right)+\frac{1}{4}\sin^{2}\xi\left(d\vartheta_{2}^{2}+\sin^{2}\vartheta_{2}\: d\varphi_{2}^{2}\right).\label{eq:metric-in-angles}\end{align}
The charges of interest can be written using these angles: writing
$J_{\varphi_{2}}=2\sqrt{2\lambda}\int dx\:\frac{\partial\mathcal{L}}{\partial\,\partial_{t}\varphi_{2}}$
etc., we have $J=2J_{\varphi_{2}}$, $Q=2J_{\varphi_{1}}$ and $J_{3}=J_{\eta}$.

\subsection{Recycled giant magnons in $CP^{3}$\label{sub:String-Magnons}}

The Hoffman--Maldacena giant magnon \cite{Hofman:2006xt} is a rigidly
rotating classical string solution in $\mathbb{R}\times S^{2}$. Writing
$S^{2}\subset\mathbb{C}^{2}$, the solution is:\begin{equation}
\vec{w}=\left(\begin{array}{l}
e^{it}\left[\cos\frac{p}{2}+i\sin\frac{p}{2}\,\tanh u\right]\\
\sin\frac{p}{2}\,\sech u\end{array}\right)=\left(\begin{array}{l}
e^{i\phi_{\mathrm{mag}}(x,t)}\sin\theta_{\mathrm{mag}}(x,t)\\
\cos\theta_{\mathrm{mag}}(x,t)\end{array}\right)\label{eq:basic-magnon}\end{equation}
where $u=\gamma(x-vt)$. The only parameter is the worldsheet velocity
$v=\cos(p/2)$. There are two ways to embed this solution into $CP^{3}$:
\begin{itemize}
\item The first class of giant magnons in $CP^{3}$ is obtained by placing
this solution into the subspace $CP^{1}=S^{2}$ defined by $z_{2}=z_{3}=0$,
or $\xi=\frac{\pi}{2}$ \cite{Gaiotto:2008cg}. With our conventions
this is a sphere is of radius $\frac{1}{2}$, and so to maintain timelike
conformal gauge the solution should have angles on this sphere of
$\vartheta_{2}=\theta_{\mathrm{mag}}(2x,2t)$ and $\varphi_{2}=\phi_{\mathrm{mag}}(2x,2t)$.
In terms of $\vec{z}$ the solution is then \begin{align}
\vec{z}(x,t) & =\left(\begin{array}{c}
e^{\frac{i}{2}\phi_{\mathrm{mag}}(2x,2t)}\sin\left(\tfrac{1}{2}\theta_{\mathrm{mag}}(2x,2t)\right)\\
0\\
0\\
e^{-\frac{i}{2}\phi_{\mathrm{mag}}(2x,2t)}\cos\left(\tfrac{1}{2}\theta_{\mathrm{mag}}(2x,2t)\right)\end{array}\right).\label{eq:CP1-magnon}\end{align}
These magnons have dispersion relation\[
\mathcal{E}(p)=\Delta-\frac{J}{2}=\sqrt{2\lambda}\sin\left(\frac{p}{2}\right).\]

\item The second class of giant magnons live in $RP^{2}$ \cite{Gaiotto:2008cg},
and can be written $\xi=\theta_{\mathrm{mag}}(x,t)$, $\varphi_{2}=2\phi_{\mathrm{mag}}(x,t)$,
$\vartheta_{1}=\vartheta_{2}=\frac{\pi}{2}$, or \cite{Grignani:2008is}\begin{equation}
\vec{z}(x,t)=\frac{1}{\sqrt{2}}\left(\begin{array}{c}
e^{i\phi_{\mathrm{mag}}(x,t)}\sin\theta_{\mathrm{mag}}(x,t)\\
\cos\theta_{\mathrm{mag}}(x,t)\\
\cos\theta_{\mathrm{mag}}(x,t)\\
e^{-i\phi_{\mathrm{mag}}(x,t)}\sin\theta_{\mathrm{mag}}(x,t)\end{array}\right)=\frac{1}{\sqrt{2}}\left(\begin{array}{c}
w_{1}\\
w_{2}\\
\bar{w}_{2}\\
\bar{w}_{1}\end{array}\right).\label{eq:RP2-magnon}\end{equation}
The dispersion relation for this class of magnons is \[
\mathcal{E}=\Delta-\frac{J}{2}=2\sqrt{2\lambda}\sin\left(\frac{p'}{2}\right)\]
where we call the velocity $v=\cos(p'/2)$ in this case because it
turns out that the momenta are related $p=2p'$.
\end{itemize}
The dyonic generalisation of the $RP^{2}$ magnon is an embedding
of Dorey's original $S^{3}$ dyonic magnon, and carries a second momentum
$Q\neq0$. Parameterising the $S^{3}$ by $\vec{w}\in\mathbb{C}^{2}$,
the embedding needed is exactly the formula \eqref{eq:RP2-magnon}
above. The resulting solution lives in $RP^{3}$ and has dispersion
relation \[
\mathcal{E}=\Delta-\frac{J}{2}=\sqrt{\frac{Q^{2}}{4}+8\lambda\sin\left(\frac{p'}{2}\right)}.\]
The third angular momentum $J_{3}$ is still zero.

All giant magnons are pieces of a closed string, and for the $CP^{1}$
magnon, like the original $S^{2}$ solution, the condition that a
set of such pieces close is $\sum_{i}p_{i}=0\mbox{ mod }2\pi$, since
$p$ is the opening angle along the equator. For the $RP^{2}$ magnon,
however, the $p'=\pi$ magnon is also a closed string, thanks to the
$\mathbb{Z}_{2}$ identification in this space, and thus $\sum_{i}p_{i}'=0\mbox{ mod }\pi$
instead. 

For more detailed discussion of these subspaces, and of others such
as $S^{2}\times S^{2},$ see \cite{Abbott:2008qd}.

\subsection{Finite-size corrections\label{sub:String-finite-J}}

For the two embeddings of $S^{2}$ giant magnons discussed above,
we can obtain finite-$J$ corrections by simply embedding the $S^{2}$
results \cite{Arutyunov:2006gs,Astolfi:2007uz} into $CP^{3}$. The
explicit calculation of these corrections was done by \cite{Grignani:2008te}
for the $RP^{2}$ case:\begin{equation}
\Delta-\frac{J}{2}=2\sqrt{2\lambda}\sin\left(\frac{p}{2}\right)\left[1-4\sin^{2}\left(\frac{p'}{2}\right)e^{-2\Delta\big/2\sqrt{2\lambda}\sin(\frac{p'}{2})}+\ldots\right]\label{eq:finite-J-RP2}\end{equation}
and by \cite{Lee:2008ui} for the $CP^{1}$ case: \begin{equation}
\Delta-\frac{J}{2}=\sqrt{2\lambda}\sin\left(\frac{p}{2}\right)\left[1-4\sin^{2}\left(\frac{p}{2}\right)e^{-2\Delta\big/\sqrt{2\lambda}\sin(\frac{p}{2})}+\ldots\right].\label{eq:finite-J-CP1}\end{equation}

For the $RP^{3}$ dyonic giant magnon, we can similarly embed the
results from $S^{3}$. These were originally computed by \cite{Hatsuda:2008gd}
(from the all-$J$ solutions of \cite{Okamura:2006zv}) and were studied
in $CP^{3}$ by \cite{Ahn:2008hj,Ahn:2008wd}. The result is \begin{equation}
\Delta-\frac{J}{2}=\sqrt{\frac{Q^{2}}{4}+8\lambda\sin^{2}\left(\frac{p'}{2}\right)}-32\lambda\cos(2\phi)\frac{1}{\mathcal{E}}\sin^{4}\left(\frac{p'}{2}\right)e^{-\Delta\mathcal{E}\big/2S}\label{eq:finite-J-RP3}\end{equation}
where we define%
\footnote{Note that $S(\frac{p'}{2})\to\frac{1}{4}\mathcal{E}^{2}$ as $Q\to0$.
When comparing to \cite{Hatsuda:2008gd}, note that $\cosh(\theta/2)=\mathcal{E}\big/2\sqrt{2\lambda}\sin(\frac{p'}{2})$.
In terms of our later notation, this $\theta$ is defined $r=e^{\theta/2}$.%
} \begin{align}
S & =\frac{Q^{2}}{16\sin^{2}(\frac{p'}{2})}+2\lambda\sin^{2}\left(\frac{p'}{2}\right).\label{eq:defn-S-string}\end{align}

\begin{figure}
\begin{centering}
\includegraphics[width=0.55\columnwidth]{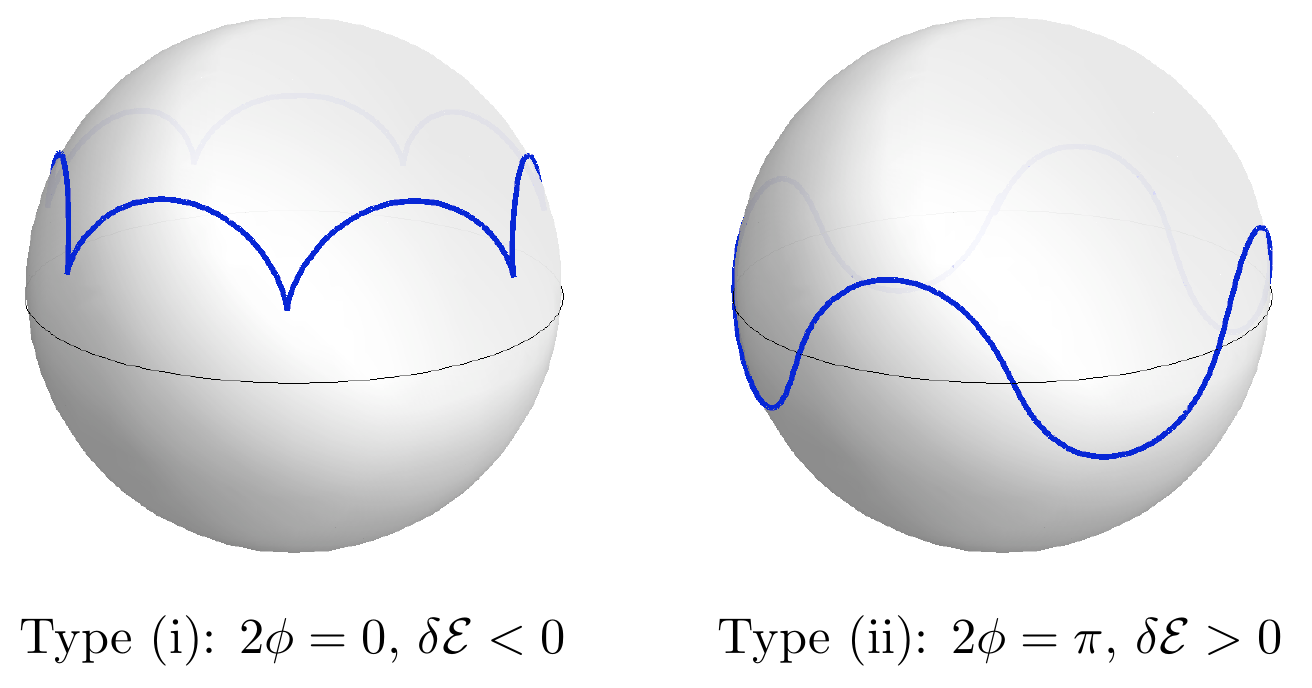}
\par\end{centering}

\caption{The two classes of finite-$J$ magnons found by \cite{Okamura:2006zv}.
\label{fig:Type-i-and-ii-strings}}

\end{figure}

It remains to discuss the factor $\cos(2\phi)$. In the paper \cite{Hatsuda:2008gd},
this is set to be $+1$ for `type (i) helical strings', and $-1$
for `type (ii)' strings. These are two kinds of finite-$J$ solutions,
which in the non-dyonic case in $S^{2}$ give a set of magnons in
which adjacent magnons either have the same or opposite orientation.
Type (i) strings thus have a cusp not touching the equator, while
type (ii) strings cross the equator at less than a right angle ---
see figure \ref{fig:Type-i-and-ii-strings}. It seems very likely
that we should interpret $2\phi$ as the angle between the two magnons'
orientation vectors. (The same factor could be included in the non-dyonic
cases \eqref{eq:finite-J-RP2} and \eqref{eq:finite-J-CP1}.)

\subsection{Dressing method solution\label{sub:Dressing-method-solution}}

There is also another kind of giant magnon, which does not exist in
$S^{5}$, recently constructed by several groups using the dressing
method \cite{Hollowood:2009tw,Kalousios:2009mp,Suzuki:2009sc}.

This method is a way of generating multi-soliton solutions above a
given vacuum in the principal chiral model, closely related to the
B\"{a}cklund transformation. (See \cite{Harnad:1983we,Sasaki:1984tp}.)
The `dressed' solution $\Psi$ is obtained from the `bare' vacuum
$\Psi_{0}$ by $\Psi=\chi(0)\Psi_{0}$, where the dressing matrix
$\chi(\lambda)$ is a function of a spectral parameter. Each independent
pole%
\footnote{Sometimes there is also an image pole at $1/\lambda_{1}$.%
} of $\chi(\lambda)$ results in one soliton, and in the cases of interest
here, its position contains the solution's parameters. 

The dressing method was first used to generate giant magnons in $S^{3}$
by \cite{Spradlin:2006wk}, where the string sigma-model was mapped
to an $SU(2)$ principal chiral model and an $SO(3)$ vector model.
In the $SU(2)$ case, a pole at $\lambda_{1}=re^{ip/2}$ produces
a dyonic giant magnon with charges\begin{align*}
\Delta-J_{1} & =\frac{\sqrt{\lambda}}{\pi}\frac{1+r^{2}}{2r}\sin\left(\frac{p}{2}\right), & J_{2} & =\frac{\sqrt{\lambda}}{\pi}\frac{1-r^{2}}{2r}\sin\left(\frac{p}{2}\right)\end{align*}
which can then be combined to give the usual dispersion relation.
(One then regards $J_{2}$ as the second parameter, instead of $r$.)
In the $SO(3)$ case, only the non-dyonic giant magnon can be obtained.

The recently constructed $CP^{3}$ solution uses the map to an $SU(4)/U(3)$
model. The position of the dressing pole $\lambda_{1}=re^{ip'/2}$
again provides two parameters, but unlike the $S^{3}$ case, there
is only one nonzero angular momentum\[
J=2\Delta-4\sqrt{2\lambda}\frac{1+r^{2}}{2r}\sin\left(\frac{p'}{2}\right).\]
(As in $RP^{2}$, we will call this momentum $p'$.) It is convenient
however to use, instead of $r$, a new parameter defined $Q_{f}=2\sqrt{2\lambda}\frac{1-r^{2}}{2r}\sin(\frac{p'}{2})$.
In terms of this, the dispersion relation becomes\[
\Delta-\frac{J}{2}=\sqrt{Q_{f}^{2}+8\lambda\sin^{2}\left(\frac{p'}{2}\right)}.\]
We have chosen the factor in front of $Q_{f}$ to make this line up
with the dispersion relation for the big giant magnon in the algebraic
curve (after setting $p=2p'$). Like this solution, the big giant
magnon is a two-parameter single-momentum solution. We discuss it
in section \ref{sub:Curve-Magnons} below.

In the basis used by \cite{Hollowood:2009tw,Kalousios:2009mp},%
\footnote{The paper \cite{Suzuki:2009sc} obtains this solution in the same
basis as we use. %
} and writing only the GKP%
\footnote{See footnote \ref{fn:GKP} about this name.%
} case $p'=\pi$, the solution is:%
\footnote{Like the $RP^{2}$ and $RP^{3}$ magnons, this forms a closed string
at $p'=\pi$.%
}\[
\vec{z}'=N\left(\begin{array}{c}
(1+r^{2})\cos(t)+\cos\left(\frac{1-3r^{2}}{1+r^{2}}t\right)+r^{2}\cos\left(\frac{3-r^{2}}{1+r^{2}}t\right)+i(1-r^{2})\sin(t)\sinh\left(\frac{4r}{1+r^{2}}x\right)\\
-(1+r^{2})\sin(t)+\sin\left(\frac{1-3r^{2}}{1+r^{2}}t\right)-r^{2}\sin\left(\frac{3-r^{2}}{1+r^{2}}t\right)-i(1-r^{2})\cos(t)\sinh\left(\frac{4r}{1+r^{2}}x\right)\\
2(1-r^{2})\left[\sin\left(\frac{1-r^{2}}{1+r^{2}}t\right)\sinh\left(\frac{2r}{1+r^{2}}x\right)-i\cos\left(\frac{1-r^{2}}{1+r^{2}}t\right)\cosh\left(\frac{2r}{1+r^{2}}x\right)\right]\\
0\end{array}\right)\]
where $N$ is a normalisation factor ensuring $\left|\vec{z}\right|^{2}=1$.
The vacuum used to derive this solution is $\vec{z}'_{\mathrm{vac}}=\left(\cos(t),\sin(t),0,0\right)$,
which carries large charge under $J'=J[\sigma_{2}\oplus1]$. The value
of this $J'$ is altered by the presence of this magnon, but all other
$J[T^{a}]$ charges vanish. Rotating the space to bring this vacuum
$\vec{z}_{\mathrm{vac}}'$ to match our $\vec{z}_{\mathrm{vac}}=\frac{1}{\sqrt{2}}\left(e^{it},0,0,e^{-it}\right)$
will also rotate the charge $J'$ into $J$, as used for the other
magnons.

In the limit $r\to1$, or $Q_{f}\to0$, the dispersion relation becomes
that of the $RP^{2}$ giant magnon. And the solution (in this basis)
becomes the embedding of the ordinary magnon \eqref{eq:basic-magnon}
given by $\vec{z}'=(\re w_{1},\im w_{1},\re w_{2},0)\in\mathbb{R}^{4}$. 

Finite-$J$ corrections to this solution are not known in the string
sigma-model (except trivially at $Q_{f}=0$, where it is the $RP^{2}$
magnon) but we compute them using the algebraic curve in section \ref{sec:Curve-Finite-J}
below.

\begin{table}
\begin{centering}
\renewcommand{\arraystretch}{2} \small 
\begin{tabular}{|l|cccc|}
\hline 
 & $\mathcal{E}=\Delta-\frac{J}{2}$ & $\delta\mathcal{E}$ (finite $J$) & $Q$ & $J_{3}$\tabularnewline
\hline
Vacuum & 0 &  &  & \tabularnewline
\hline
$CP^{1}$ giant magnon & $\sqrt{2\lambda}\sin(\frac{p}{2})$ & $-4\mathcal{E}\sin^{2}(\frac{p}{2})e^{-2\Delta/\mathcal{E}}$ & 0 & 0\tabularnewline
Dyonic version in $CP^{2}$ \newstar$\vphantom{\dfrac{A_{B}^{A}}{B_{B}^{A}}}$ & $\sqrt{\frac{Q^{2}}{4}+2\lambda}$ when $p=\pi$ & (Use `small' curve, \eqref{eq:correction-small}) & $Q$ & $\pm Q$\tabularnewline
\hline
$RP^{2}$ giant magnon & $2\sqrt{2\lambda}\sin(\frac{p'}{2})$ & $-4\mathcal{E}\sin^{2}(\frac{p'}{2})e^{-2\Delta/\mathcal{E}}$ & 0 & 0\tabularnewline
Dyonic version in $RP^{3}$ & $\sqrt{\frac{Q^{2}}{4}+8\lambda\sin^{2}(\frac{p'}{2})}$ & Like $S^{5}$ result, \eqref{eq:finite-J-RP3} & $Q$ & 0\tabularnewline
HM/KSV/S dressed solution$\vphantom{\dfrac{A_{B}^{A}}{B_{B}^{A}}}$ & $\sqrt{Q_{f}^{2}+8\lambda\sin^{2}(\frac{p'}{2})}$ & (Use `big' curve, \eqref{eq:correction-big}) & $0$ & 0\tabularnewline
\hline
\end{tabular}
\par\end{centering}

\caption{Summary of giant magnons in the string sigma-model.\label{tab:String-Table}
The dressed solution of \cite{Hollowood:2009tw,Kalousios:2009mp,Suzuki:2009sc}
also lives in $CP^{2}$. (The $RP^{2}$ solution has often been called
$SU(2)\times SU(2)$, `big' and $S^{2}\times S^{2}$ in the literature.)
To match the curves we want $p'=p/2$.}

\end{table}

\subsubsection*{Note added\label{sub:Note-added}}

The paper \cite{Kalousios:2009mp} also finds a second solution by
dressing, equation (4.14).%
\footnote{We thank Chrysostomos Kalousios for drawing our attention to this
solution.%
} This solution has the same angular momentum $J$ as the big giant
magnon, but lives in $CP^{1}$. It is in fact just an embedding of
the bound state solution (5.14) of \cite{Spradlin:2006wk}, which
in that paper is written using parameter $q$ instead of $r=e^{q/2}$.
This bound state is an analytic continuation of a scattering state
of two HM magnons, or in this case, of two $CP^{1}$ magnons.

\section{Dyonic generalisation of the $CP^{1}$ magnon\label{sec:CP1-Dyonic}}

Consider the subspace $CP^{2}$ obtained by fixing $z_{3}=0$, or
in terms of the angles, $\theta_{1}=0$ and $\eta=0$, leaving \begin{equation}
\vec{z}=\left(\begin{array}{l}
\sin\xi\:\cos(\vartheta_{2}/2)\: e^{i\varphi_{2}/2}\\
\cos\xi\: e^{i\varphi_{1}/2}\\
0\\
\sin\xi\:\sin(\vartheta_{2}/2)\: e^{-i\varphi_{2}/2}\end{array}\right).\label{eq:CP2-subspace-angles}\end{equation}
The metric for this subspace can be written\[
ds^{2}=\frac{1}{4}\sin^{2}\xi\left[d\vartheta_{2}^{2}+\sin^{2}\vartheta_{2}\: d\varphi_{2}^{2}\;+\cos^{2}\xi\left(d\varphi_{1}-\cos\vartheta_{2}\: d\varphi_{2}\right)^{2}\right]+d\xi^{2}.\]
At $\xi=\frac{\pi}{2}$ the space is $CP^{1}$, described by $\vartheta_{2}$
and $\varphi_{2}$ only, but away from this value there is a second
isometry direction $d\varphi_{1}$. It was proposed in \cite{Abbott:2008qd}
that the dyonic generalisation of the $CP^{1}$ magnon might have
momentum along this direction, but to do so, it must in addition have
$\xi\neq\frac{\pi}{2}$ except at the endpoints of the string, at
$x=\pm\infty$, where it must touch the same equator as the $CP^{1}$
solution. 

We have not yet been able to find the full solution, but can find
a GKP-like dyonic solution (i.e. a $p=\pi$ magnon%
\footnote{GKP \cite{Gubser:2002tv} studied rotating folded strings, which at
$J=\infty$ are the $p=\pi$ case of the HM magnon. \cite{Hofman:2006xt}
Two-spin folded string solutions were studied by F\&T, \cite{Frolov:2003qc}
and are the $p=\pi$ case of Dorey's dyonic giant magnon. \cite{Dorey:2006dq,Chen:2006gea}\label{fn:GKP}%
}) using the ansatz:\begin{align*}
\varphi_{2} & =2t, & \varphi_{1} & =-2\omega t,\\
\cos\vartheta_{2} & =\sech\left(\sqrt{1-\omega^{2}}2x\right), & \xi & =\frac{\pi}{2}-e(x).\end{align*}
This amounts to assuming that the back-reaction on the original solution
in $\vartheta_{2},\varphi_{2}$ created by giving it new momentum
along $\varphi_{1}$ is exactly as for the $S^{3}$ dyonic solution,
but unlike the $S^{3}$ case, there is one extra function $e(x)$. 

With this ansatz the equations of motion for $\varphi_{1}$ and $\varphi_{2}$,
and the second Virasoro constraint, are already solved. The equation
of motion for $\vartheta_{2}$ can be written \[
\partial_{x}\left(\cos^{2}e(x)\,\sech X\right)=-2\frac{\cos^{2}e(x)}{\sqrt{1-\omega^{2}}}\tanh X\left\{ \sech X\cos^{2}e(x)+\omega\sin^{2}e(x)\right\} ,\]
 where $X=\sqrt{1-\omega^{2}}2x$. Using a change of variables $y(x)=\ln\left(\cos^{2}e(x)\right)$,
this equation can be written as \begin{equation}
y'(x)=e^{y(x)}f(x)+g(x)\,,\label{eq:form-of-the-diff-eq-for-e}\end{equation}
 where\begin{align*}
f(x) & =\frac{2}{\sqrt{1-\omega^{2}}}\big(\omega-\sech X\big)\sinh X,\\
g(x) & =-f(x)-\frac{2\omega^{2}}{\sqrt{1-\omega^{2}}}\tanh X.\end{align*}
The equation \eqref{eq:form-of-the-diff-eq-for-e} has solutions of
the form $y(x)=-\ln\left(-F(x)\right)+G(x)$, where $G'(x)=g(x)$
and $F'(x)=f(x)e^{G(x)}$. After some algebra, we find the following
form for the solution: \begin{equation}
\cos^{2}\left(e(x)\right)=\sin^{2}\xi=\frac{1}{1+\omega\cos\vartheta_{2}}=\frac{1}{1+\omega\sech\left(\sqrt{1-\omega^{2}}2x\right)}\label{eq:our-CP2-soln-e}\end{equation}
where $\omega\geq0$.

\begin{figure}
\begin{centering}
\includegraphics[width=0.8\columnwidth]{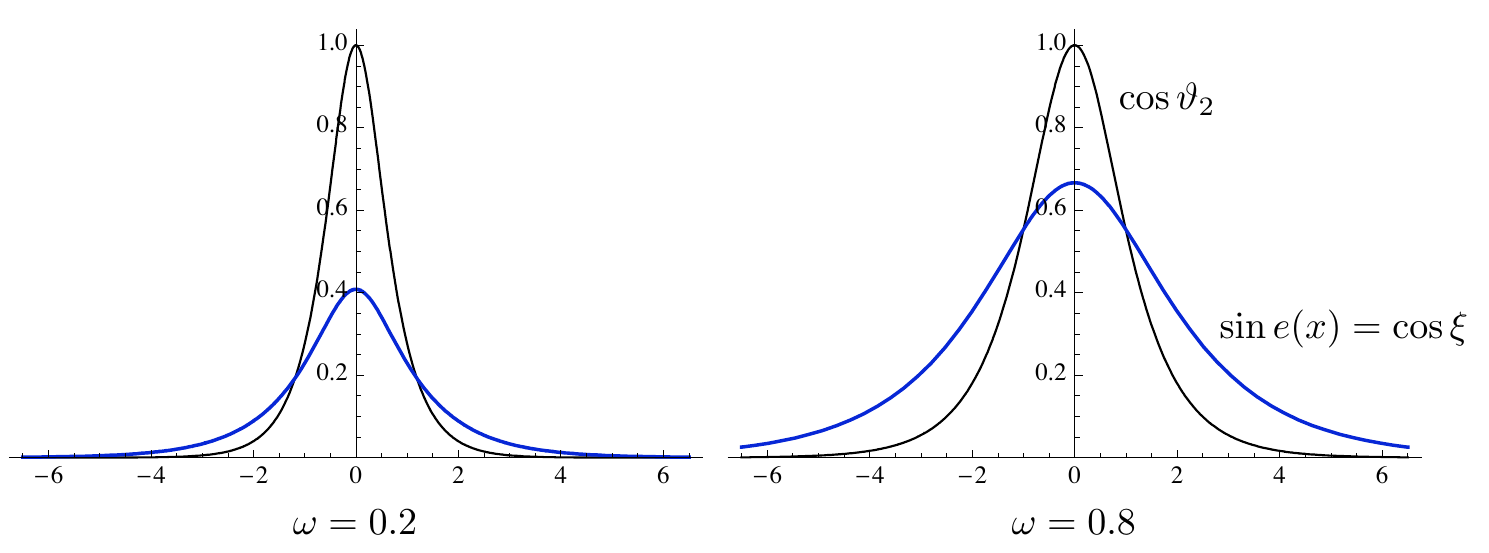}
\par\end{centering}

\caption{Profiles of the $CP^{2}$ solution. $\cos\vartheta_{2}$ is the same
as for Dorey's $S^{3}$ dyonic giant magnon, but the solution also
spreads away from $\xi=\frac{\pi}{2}$ as we increase $\omega$. This
is shown for $\omega=0.2$ and $0.8$.}

\end{figure}

Calculating charges $J$ and $Q$ for this solution, we find \begin{align*}
\Delta-\frac{J}{2} & =\sqrt{2\lambda}\frac{1}{\sqrt{1-\omega^{2}}}, & \frac{Q}{2} & =-\sqrt{2\lambda}\frac{\omega}{\sqrt{1-\omega^{2}}}\end{align*}
 and therefore the dispersion relation is \[
\Delta-\frac{J}{2}=\sqrt{\frac{Q^{2}}{4}+2\lambda\:}.\]
We conjecture that for the general case (allowing $p\neq\pi$) the
dispersion relation is\[
\mathcal{E}=\Delta-\frac{J}{2}=\sqrt{\frac{Q^{2}}{4}+2\lambda\sin\left(\frac{p}{2}\right)}\]
matching the one for the `small giant magon' in the algebraic curve. 

Unlike the $RP^{3}$ dyonic magnon, this one is charged not only under
$Q$ but also under $J_{3}$, with $J_{3}=Q$. There is a second $CP^{2}$
solution, in the subspace with $z_{2}=0$ instead of $z_{3}=0$, which
has $J_{3}=-Q$ but is otherwise similar. In the limit $\omega\to0$
both kinds become the same $CP^{1}$ solution. All of these properties
match those of the two kinds of small giant magnons in the algebraic
curve perfectly.

We summarise all the properties of the various string solutions in
table \ref{tab:String-Table}.

\section{The algebraic curve for $AdS_{4}\times CP^{3}$\label{sec:The-algebraic-curve}}

The string equations of motion can also be studied using the formalism
of algebraic curves. This has been a fruitful approach in $AdS_{5}\times S^{5}$
\cite{Kazakov:2004qf,Kazakov:2004nh,Beisert:2004ag,SchaferNameki:2004ik,Beisert:2005bm,Beisert:2005di}.
The $AdS_{4}\times CP^{3}$ case at hand was originally studied by
Gromov and Vieira \cite{Gromov:2008bz}. We start this section by
a brief review of the construction of the algebraic curve and its
most important properties.

\subsection{From target space to quasi-momenta}

Begin by defining the connection $j=j_{AdS}\oplus j_{CP}$, where
\cite{Stefanski:2008ik,Arutyunov:2008if} \begin{align*}
(j_{AdS})_{ij,\mu} & =2\left(y_{i}\partial_{\mu}y_{j}-(\partial_{\mu}y_{i})y_{j}\right)\\
(j_{CP})_{ij,\mu} & =2\left(z_{i}D_{\mu}z_{j}-(D_{\mu}z_{i})z_{j}\right)\end{align*}
By construction, the connection $j$ is flat\[
dj+j\wedge j=0\,.\]
The sigma-model action can be written in terms of $j$ as\[
S=-\frac{g}{8}\int d\sigma\, d\tau\:\STr\, j^{2}\,,\]
leading to the equations of motion\begin{equation}
d*j=0\,.\label{eq:eom-j}\end{equation}

The Lax connection is now given by\[
J(x)=\frac{1}{1-x^{2}}j+\frac{x}{1-x^{2}}*j\]
where the new complex variable $x$ is the spectral parameter.%
\footnote{In this section we use $x$ to be the spectral parameter, and $\sigma,\tau$
to be the worldsheet co-ordinates which we called $x,t$ before.%
} $J(x)$ is a flat connection, for any $x$, provided $j$ is flat
and satisfies \eqref{eq:eom-j}.

Using $J(x)$ we define the monodromy matrix\[
\Omega(x)=P\, e^{\int d\sigma J_{\sigma}(x)}\,.\]
Since the connection is flat, the eigenvalues of $\Omega$ are independent
of $\tau$. We write these eigenvalues as $e^{i\tilde{p}_{1}},e^{i\tilde{p}_{2}},e^{i\tilde{p}_{3}},e^{i\tilde{p}_{4}}$
for the $CP^{3}$ part, and $e^{i\hat{p}_{1}},e^{i\hat{p}_{2}},e^{i\hat{p}_{3}},e^{i\hat{p}_{4}}$
for the $AdS$ part, and refer to the functions $\tilde{p}_{i}$ and
$\hat{p}_{i}$ as `quasi-momenta'. The continuity in the complex plane
of the function $\mathrm{eig(\Omega(x))}$ demands that when a branch
cut $C_{ij}$ connects sheets $i$ and $j$, we must have $p_{i}^{+}-p_{j}^{-}=2\pi n$
when $x\in C_{ij}$.

At large $x$ the monodromy matrix is\[
\Omega(x)=1+\frac{1}{x}\int d\sigma j_{\tau}+\ldots\]
and thus the asymptotic behaviour of the quasi-momenta contains information
about the charges. (The exact relation is \eqref{eq:asympt-curve-charge}
below.)

The quasi-momenta $\tilde{p}_{i}$ and $\hat{p}_{i}$ describe the
bosonic sector of type IIA string theory on $AdS_{4}\times CP^{3}$.
While this is all we will need in this paper, it will be convenient
to work in a formalism with explicit $OSp(2,2|6)$ symmetry. To do
this we define ten new quasi-momenta $q_{i}$ as \cite{Gromov:2008bz}
\[
\{q_{1},q_{2},q_{3},q_{4},q_{5}\}=\frac{1}{2}\{\hat{p}_{1}+\hat{p}_{2},\hat{p}_{1}-\hat{p}_{2},\tilde{p}_{1}+\tilde{p}_{2},-\tilde{p}_{2}-\tilde{p}_{4},\tilde{p}_{1}+\tilde{p}_{4}\}\]
and\[
\{q_{6},q_{7},q_{8},q_{9},q_{10}\}=\{-q_{5},-q_{4},-q_{3},-q_{2},-q_{1}\}\,.\]
The functions $q_{i}$ now define a ten-sheeted Riemann surface.

\subsection{Relations and charges}

The ten quasi-momenta $q_{i}$. must obey the following relations:
\cite{Gromov:2008bz}
\begin{enumerate}
\item Only five are independent: $\{q_{6},q_{7},q_{8},q_{9},q_{10}\}=\{-q_{5},-q_{4},-q_{3},-q_{2},-q_{1}\}$.
\item Square-root branch cut condition: $q_{i}^{+}(x)-q_{j}^{-}(x)=2\pi n_{ij}$,
$x\in C_{ij}$.
\item Synchronised poles: the residues at $x=\pm1$ are the same $\alpha_{\pm}/2$
for $q_{1},q_{2},q_{3},q_{4}$, while $q_{5}$ does not have a pole
there.
\item Inversion symmetry: $q_{1}(\frac{1}{x})=-q_{2}(x)$, $q_{3}(\frac{1}{x})=2\pi m-q_{4}(x)$
and $q_{5}(\frac{1}{x})=q_{5}(x)$. This $m\in\mathbb{Z}$ gives the
momentum: $p=2\pi m$.
\item Asymptotic behaviour as $x\to\infty$: \begin{equation}
\left(\begin{array}{c}
q_{1}\\
q_{2}\\
q_{3}\\
q_{4}\\
q_{5}\end{array}\right)=\frac{1}{2gx}\left(\begin{array}{c}
\Delta+S\\
\Delta-S\\
L-M_{r}\\
L+M_{r}-M_{u}-M_{v}\\
M_{v}-M_{u}\end{array}\right)+o(\frac{1}{x^{2}})=\frac{1}{2gx}\left(\begin{array}{c}
\Delta+S\\
\Delta-S\\
J_{1}\\
J_{2}\\
J_{3}\end{array}\right)+\ldots\label{eq:asympt-curve-charge}\end{equation}
where $\lambda=8g^{2}$ (i.e. $4g=\sqrt{2\lambda}$). 
\end{enumerate}
The ansatz used by \cite{Lukowski:2008eq}, for solutions mostly in
$CP^{3}$, is the following:\begin{equation}
\begin{array}{ccccc}
q_{1}(x)= & \negmedspace\negmedspace\dfrac{\alpha x}{x^{2}-1} &  &  & \tablespacer\\
q_{2}(x)= & \negmedspace\negmedspace\dfrac{\alpha x}{x^{2}-1} &  &  & \tablespacer\\
q_{3}(x)= & \negmedspace\negmedspace\dfrac{\alpha x}{x^{2}-1} & \negmedspace\negmedspace+G_{u}(0)-G_{u}(\frac{1}{x})\tablespacer & +G_{v}(0)-G_{v}(\frac{1}{x})\quad & \negmedspace\negmedspace\negmedspace\negmedspace\negmedspace\negmedspace+G_{r}(x)-G_{r}(0)+G_{r}(\frac{1}{x})\\
q_{4}(x)= & \negmedspace\negmedspace\dfrac{\alpha x}{x^{2}-1} & \negmedspace\negmedspace+G_{u}(x)\tablespacer & +G_{v}(x)\quad & \negmedspace\negmedspace\negmedspace\negmedspace\negmedspace\negmedspace-G_{r}(x)+G_{r}(0)-G_{r}(\frac{1}{x})\\
q_{5}(x)= &  & \negmedspace\negmedspace\negmedspace\negmedspace G_{u}(x)-G_{u}(0)+G_{u}(\frac{1}{x}) & \negmedspace\negmedspace-G_{v}(x)+G_{v}(0)-G_{v}(\frac{1}{x})\end{array}\label{eq:q12345-ansatz}\end{equation}
From $q_{1}$ and $q_{2}$ we can read off $S=0$ (zero $AdS$ angular
momentum) and $\alpha=\Delta/2g$. 

The functions $G_{u},G_{v},G_{r}$ control the $CP^{3}$ part of the
curve $q_{3},q_{4},q_{5}$ and so, asymptotically, the $SU(4)$ excitation
numbers $M_{u},M_{v},M_{r}$. Their values at $x=0$ control the momentum%
\footnote{For a closed string $m\in\mathbb{Z}$, however, we want to consider
a single giant magnon which in general is not a closed string. Hence
we will relax this condition and consider general $p$. To get a physical
state this momentum condition should be imposed. This can be done
by considering multi-magnon states \cite{Minahan:2008re,Lukowski:2008eq}.%
} \begin{equation}
p=2\pi m=q_{3}\left(\frac{1}{x}\right)+q_{4}(x).\label{eq:defn-p-from-q3q4}\end{equation}
The Dynkin labels of $SU(4)$ are related to the excitation numbers
by \[
\left[\begin{array}{c}
p_{1}\\
q\\
p_{2}\end{array}\right]=\left[\begin{array}{c}
L-2M_{u}+M_{r}\\
M_{u}+M_{v}-2M_{r}\\
L-2M_{v}+M_{r}\end{array}\right]\in\mathbb{Z}_{\geq0}^{3}\]
These can be combined into the $SO(6)$ charges: \[
\left(\begin{array}{c}
J_{1}\\
J_{2}\\
J_{3}\end{array}\right)=\left(\begin{array}{c}
q+(p_{1}+p_{2})/2\\
(p_{1}+p_{2})/2\\
(p_{2}-p_{1})/2\end{array}\right)=\left(\begin{array}{c}
L-M_{r}\\
L+M_{r}-M_{u}-M_{v}\\
M_{u}-M_{v}\end{array}\right)\]
which are in turn combined into the magnons' major and minor charges:\begin{align*}
J & =J_{1}+J_{2}\;=2L-M_{u}-M_{v}\quad\:=p_{1}+q+p_{2}\\
Q & =J_{1}-J_{2}\;=M_{u}+M_{v}-2M_{r}\;=q\end{align*}

\subsection{Giant magnons in the curve\label{sub:Curve-Magnons}}

Giant magnons were first studied using the algebraic curve in \cite{Minahan:2006bd},
where it was shown that they correspond to logarithmic cuts (see also
\cite{Vicedo:2007rp}). For the case of $CP^{3}$, two different kinds
of giant magnons were given by \cite{Shenderovich:2008bs}, who named
them `small' and `big'. These can be constructed by setting some of
the resolvents in the above ansatz to \begin{equation}
G_{\mathrm{mag}}(x)=-i\log\left(\frac{x-X^{+}}{x-X^{-}}\right)\label{eq:defn-Gmag}\end{equation}
where $X^{-}$ is the complex conjugate of $X^{+}$. 
\begin{itemize}
\item The first kind is the `small giant magnon' with \[
G_{v}(x)=G_{\mathrm{mag}}(x),\qquad G_{u}=G_{r}=0.\]
The charges read off from this curve are: \begin{align}
p & =-i\log\frac{X^{+}}{X^{-}}, & Q & =-i2g\left(X^{+}-X^{-}+\frac{1}{X^{+}}-\frac{1}{X^{-}}\right),\nonumber \\
J & =2\Delta+i2g\left(X^{+}-X^{-}-\frac{1}{X^{+}}+\frac{1}{X^{-}}\right), & J_{3} & =Q.\label{eq:asympt-small}\end{align}
These can be put together to give the dispersion relation \[
\Delta-\frac{J}{2}=\sqrt{\frac{Q^{2}}{4}+16g^{2}\sin^{2}\left(\frac{p}{2}\right)}.\]

\item We can make another kind of small magnon with $G_{u}$ instead of
$G_{v}$. The only change is in the sign of $J_{3}=-Q$.
\item Then there is the `big giant magnon', which has \[
G_{u}(x)=G_{v}(x)=G_{r}(x)=G_{\mathrm{mag}}(x)\]
from which we obtain the charges \begin{align}
p & =-2i\log\frac{X^{+}}{X^{-}}, & Q & =0,\nonumber \\
J & =2\Delta+i4g\left(X^{+}-X^{-}-\frac{1}{X^{+}}+\frac{1}{X^{-}}\right), & J_{3} & =0.\label{eq:asympt-big}\end{align}
This is the curve used by \cite{Shenderovich:2008bs}, and to get
the dispersion relation, we must use not the total $Q$ but rather
$Q_{u}$, the contribution from just the $u$ part (which is cancelled
by the $v$ part in the full solution). This is the same function
of $X^{\pm}$ as for the small giant magnon \eqref{eq:asympt-small}
above. The result is \[
\Delta-\frac{J}{2}=\sqrt{Q_{u}^{2}+64g^{2}\sin^{2}\left(\frac{p}{4}\right)}.\]
For this solution, $\mathcal{E}=\Delta-J/2$ is a function of two
parameters, $Q_{u}$ and $p$, but $Q_{u}$ is not an asymptotic charge
of the full solution. Unlike the ordinary dyonic giant magnons, which
are two-parameter two-momentum solutions, here there is only one angular
momentum.
\end{itemize}
Finally, we can also put one small magnon into each sector, $G_{v}(x)=G_{u}(x)=G_{\mathrm{mag}}(x)$
but with $G_{r}=0$. For each of the charges (including both $\Delta$
and $p$) we obtain the sum of those of each of the constituent small
giant magnons, and write $Q=Q_{u}+Q_{v}$ etc. Thus we get dispersion
relation \begin{align*}
\Delta-\frac{J}{2} & =\sqrt{\frac{Q_{u}^{2}}{4}+16g^{2}\sin^{2}\left(\frac{p_{u}}{2}\right)}+\sqrt{\frac{Q_{v}^{2}}{4}+16g^{2}\sin^{2}\left(\frac{p_{v}}{2}\right)}\\
 & =\sqrt{\frac{Q^{2}}{4}+64g^{2}\sin^{2}\left(\frac{p}{4}\right)}.\end{align*}
If we were to write this in terms of the momentum $p_{u}$ of one
constituent magnon, rather than the total $p$, then we would have
$\sin^{2}(p_{u}/2)$, as in \cite{Lukowski:2008eq}. Note that this
solution has total $J_{3}=0$ (like the big magnon).

We summarise all of these properties, and more, in table \ref{tab:Curve-Table}.

\begin{table}
\begin{centering}
\renewcommand{\arraystretch}{1.5} \small 
\begin{tabular}{|lcccccc|}
\hline 
 & $M_{u}$, $M_{v}$, $M_{r}$  & $[p_{1},q,p_{2}]$ & $\mathcal{E}=\Delta-\frac{J}{2}$ & $\negmedspace\negmedspace\delta\mathcal{E}$ (finite $J$) \newstar & $Q$  & $J_{3}$\tabularnewline
\hline 
Vacuum \hspace{-10cm} &  &  &  &  &  & \tabularnewline
\hspace{5mm} & 0, 0, 0 & $[L,0,L]$ & 0 & --- & 0 & 0\tabularnewline
\hline
Small giant magnon \hspace{-10cm} &  &  &  &  &  & \tabularnewline
$\vphantom{\dfrac{A_{B}^{A}}{B_{B}^{A}}}$ & 1, 0, 0 & $[L-2,1,L]$ & $\sqrt{2\lambda}\sin(\frac{p}{2})$ & $-4\mathcal{E}\sin^{2}\left(\frac{p}{2}\right)e^{-2\Delta/\mathcal{E}}$  & 1 & 1\tabularnewline
$\vphantom{\dfrac{A_{B}^{A}}{B_{B}^{A}}}$ & Q, 0, 0 & $[L-2Q,Q,L]$ & $\sqrt{\frac{Q^{2}}{4}+2\lambda\sin^{2}(\frac{p}{2})}$ & $\propto Q/\mathcal{E}\sqrt{S}$, see \eqref{eq:correction-small} & $Q$ & $Q$\tabularnewline
... and similar with $u\leftrightarrow v$: \hspace{-10cm} &  &  &  &  &  & \tabularnewline
$\vphantom{\dfrac{A_{B}^{A}}{B_{B}^{A}}}$ & 0, Q, 0 & $[L,Q,L-2Q]$ & (same) & (same) & $Q$ & $-Q$\tabularnewline
\hline
Big giant magnon \hspace{-10cm} &  &  &  &  &  & \tabularnewline
$\vphantom{\dfrac{A_{B}^{A}}{B_{B}^{A}}}$ & 1, 1, 1 & $[L-1,0,L-1]$ & $2\sqrt{2\lambda}\sin(\frac{p}{4})$ & $-4\mathcal{E}\sin^{2}\left(\frac{p}{4}\right)e^{-2\Delta/\mathcal{E}}$  & 0 & 0\tabularnewline
$\vphantom{\dfrac{A_{B}^{A}}{B_{B}^{A}}}$ & $Q_{u}$,$Q_{u}$,$Q_{u}$ & $[L-Q_{u},0,L-Q_{u}]$ & $\sqrt{Q_{u}^{2}+8\lambda\sin^{2}(\frac{p}{4})}$ & $\propto S/\mathcal{E}Q_{u}^{2}$, see \eqref{eq:correction-big} & 0 & 0\tabularnewline
\hline
Pair of small giant magnons \hspace{-10cm} &  &  &  &  &  & \tabularnewline
$\vphantom{\dfrac{A_{B}^{A}}{B_{B}^{A}}}$ & 1, 1, 0 & $[L-2,2,L-2]$ & $2\sqrt{2\lambda}\sin(\frac{p}{4})$ & $-4\mathcal{E}\sin^{2}(\frac{p}{4})e^{-2\Delta/\mathcal{E}}$ & 2 & 0\tabularnewline
$\vphantom{\dfrac{A_{B}^{A}}{B_{B}^{A}}}$ & $\frac{Q}{2}$, $\frac{Q}{2}$, 0 & $[L-Q,Q,L-Q]$ & $\sqrt{\frac{Q^{2}}{4}+8\lambda\sin^{2}(\frac{p}{4})}$  & Like $S^{5}$ case, see \eqref{eq:correction-pair} & $Q$ & 0\tabularnewline
\hline
\end{tabular}
\par\end{centering}

\caption{Summary of giant magnons in the algebraic curve.\label{tab:Curve-Table}
In each case we list dyonic (or `dyonic') solutions, meaning $Q\sim\sqrt{\lambda}$,
below the non-dyonic case. We write these using $\lambda$ rather
than $g$ for comparison with the string sigma-model results on page
\pageref{tab:String-Table}; the relation is $\sqrt{2\lambda}=4g$.
(Note that the AFZ-like result for the pair of small magnons is not
new to this paper, it was found by \cite{Lukowski:2008eq}.)}

\end{table}

\subsection{Coalescence of non-dyonic solutions\label{sub:Coalescence-of-non-dyonic}}

Notice that in the non-dyonic limit $Q\ll g$, and $Q_{u}\ll g$,
the dispersion relations for the pair of small magnons and the big
magnon agree. This is not limited to just the dispersion relation:
in this limit, $X^{\pm}=e^{\pm ip/4}$ (in both cases) and thus we
have \begin{equation}
G_{\mathrm{mag}}(x)-G_{\mathrm{mag}}(0)+G_{\mathrm{mag}}\left(\frac{1}{x}\right)=0.\label{eq:Gx+G1/x-cancellation}\end{equation}
Looking at the ansatz \eqref{eq:q12345-ansatz}, this is equivalent
to setting $G_{r}=0$. Thus the big giant magnon becomes the same
algebraic curve as the pair of small magnons, in this limit.

For the small giant magnon, the same identity implies that $q_{5}=0$
in the non-dyonic limit. This removes the difference between curves
for the $u$ and $v$ small giant magnons.

\section{Finite-size corrections in the curve\label{sec:Curve-Finite-J}}

These corrections were studied by \cite{Lukowski:2008eq}, where the
basic technique is to replace $G_{\mathrm{mag}}(x)$ with the resolvent
\begin{equation}
G_{\mathrm{finite}}(x)=-2i\log\left(\frac{\sqrt{x-X^{+}}+\sqrt{x-Y^{+}}}{\sqrt{x-X^{-}}+\sqrt{x-Y^{-}}}\right)\label{eq:defn-Gfinite}\end{equation}
where $Y^{\pm}$ are points shifted by some small amount $\delta\ll1$
away from $X^{\pm}$:%
\footnote{Note that this is a different choice of $\phi$ to that used in \cite{Minahan:2008re,Sax:2008in,Lukowski:2008eq}.
It is chosen to separate $\phi$, which gives the orientation factor
$\cos(2\phi)$ in $\delta\mathcal{E}$, from the phase of $X^{\pm}$,
which is sometimes $p/2$ and sometimes $p/4$. %
} \begin{equation}
Y^{\pm}=X^{\pm}\left(1\pm i\delta e^{\pm i\phi}\right)\label{eq:denf-Ypm}\end{equation}
 When $\delta=0$ this new $G_{\mathrm{finite}}(x)$ clearly reduces
to the infinite-size magnon resolvent \eqref{eq:defn-Gmag}. This
form of resolvent was found based on work \cite{Minahan:2008re},
and finite-$J$ corrections to the $S^{2}$ magnon were computed using
this in \cite{Sax:2008in}.

\subsection{Finite-size small giant magnon}

The first example we study is the magnon created by setting $G_{u}(x)=G_{\mathrm{finite}}(x)$
in the general ansatz \eqref{eq:q12345-ansatz}, with $G_{v}=G_{r}=0$.
This one we discuss in the most detail, as subsequent examples are
similar. We write \[
X^{\pm}=r\, e^{ip_{0}/2}\]
in terms of which $p=p_{0}+\delta p_{(1)}+\delta^{2}p_{(2)}+o(\delta^{3})$
and\begin{align}
\mathcal{E}=\Delta-\frac{J}{2} & =4g\frac{r^{2}+1}{2r}\sin\left(\frac{p}{2}\right)-\frac{\delta}{2}J_{(1)}-\frac{\delta^{2}}{2}J_{(2)}+o(\delta^{3})\nonumber \\
Q & =8g\frac{r^{2}-1}{2r}\sin\left(\frac{p}{2}\right)+\delta\, Q_{(1)}+\delta^{2}Q_{(2)}+o(\delta^{3})\label{eq:Q-small-dyonic}\end{align}
We give formulae for these expansions in the appendix. From the full
asymptotic charges, we can calculate the energy correction in terms
of $\delta$. The first nonzero contribution is at order $\delta^{2}$:\begin{align}
\delta\mathcal{E} & =\left(\Delta-\frac{J}{2}\right)-\sqrt{\frac{Q^{2}}{4}+16g^{2}\sin^{2}\left(\frac{p}{2}\right)}\nonumber \\
 & =-\delta^{2}\frac{g}{4}\cos(2\phi)\frac{2r}{1+r^{2}}\sin\left(\frac{p}{2}\right)+o(\delta^{3})\label{eq:small-deltaE-in-delta}\end{align}

\begin{figure}
\begin{centering}
\includegraphics[width=0.6\columnwidth]{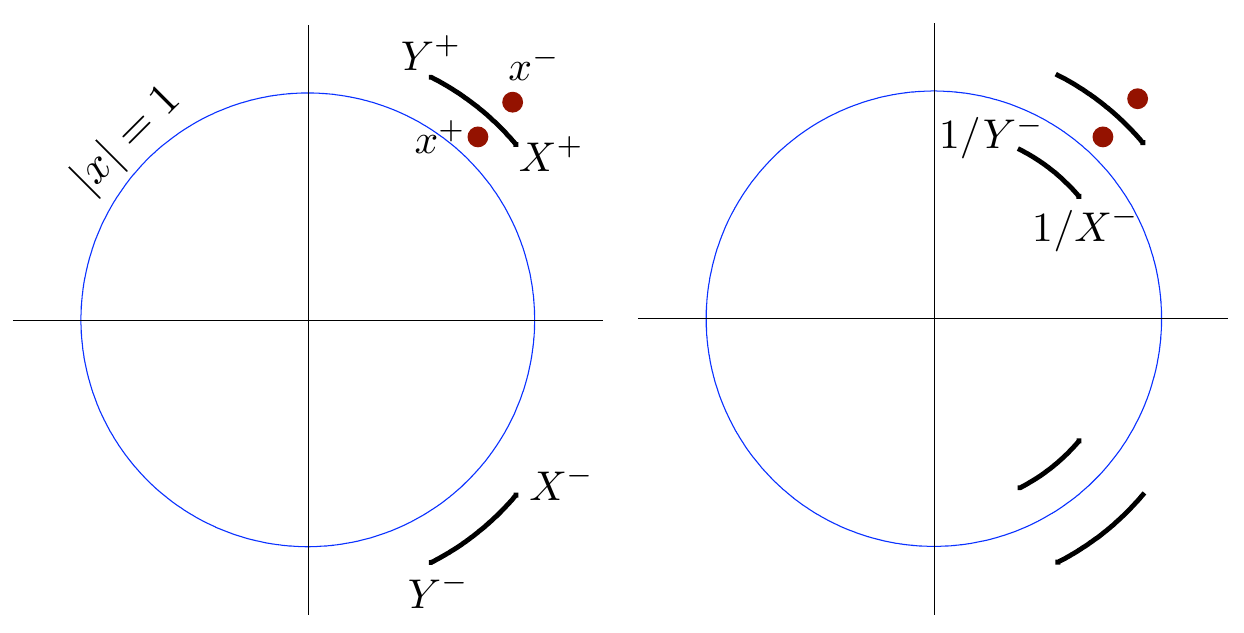}
\par\end{centering}

\caption{Branch cuts \& evaluation points. On the left, the situation for the
pair of small magnons, where only the cuts for $G_{\mathrm{finite}}(x)$
appear. The evaluation points $x^{\pm}$ straddle the cut from $X^{+}$
to $Y^{+}$, which is at radius $\left|x\right|=r$. On the right,
the cuts in $G_{\mathrm{finite}}(1/x)$ are drawn too, which is the
situation encountered in the `small' and `big' magnons. The evaluation
points are on the same side of the cut from $1/X^{+}$ to $1/Y^{+}$,
and remain so even when we take the non-dyonic limit $r\to1$. (These
are drawn for $\phi=0$.)\label{fig:Branch-cuts-Xpm}}

\end{figure}

The function $G_{\mathrm{finite}}(x)$ has a square-root branch cut
from $X^{+}$ to $Y^{+}$, which in the curve \eqref{eq:q12345-ansatz}
we choose to make connect sheets $q_{4}$ and $q_{6}=-q_{5}$. We
can then fix $\delta$ using the branch cut condition:\begin{align*}
2\pi n & =q_{4}(x^{+})-q_{6}(x^{-})\\
 & =\dfrac{2\alpha x}{x^{2}-1}+G_{\mathrm{finite}}^{+}(X^{+})+G_{\mathrm{finite}}^{-}(X^{+})-G_{\mathrm{finite}}(0)+G_{\mathrm{finite}}\left(\frac{1}{X^{+}}\right)\end{align*}
The superscript $G^{-}$ is to indicate that this term is evaluated
on the other side of the cut from the others (and thus has the opposite
sign between the terms of the numerator inside $G$). After taking
this account we may take both evaluation points $x^{\pm}$ to be at
$x=X^{+}$. Figure \ref{fig:Branch-cuts-Xpm} shows the cuts and the
points used. The result is\[
\delta=\frac{8i\: e^{-ip/4}e^{i\pi n}e^{-i\phi}\sqrt{r^{2}-1}\:\sin(\frac{p}{2})}{\sqrt{e^{-ip/2}-r^{2}e^{ip/2}}}\exp\left(\frac{i\Delta r/4g}{e^{-ip/2}-r^{2}e^{ip/2}}\right)=e^{i\psi}\left|\delta\right|\]
 In order to have a real energy correction, we demand that $\delta$
be real. We then find the correction to be \begin{align}
\delta\mathcal{E} & =-32g\cos(2\phi)\frac{r^{2}-1}{r^{2}+1}\frac{\sin^{3}\left(\frac{p}{2}\right)}{\sqrt{r^{2}+\frac{1}{r^{2}}-2\cos(p)}}e^{-\Delta\mathcal{E}\big/S(\frac{p}{2})}+o(\delta^{3})\nonumber \\
 & =-32g^{2}\cos(2\phi)\frac{Q}{\mathcal{E}\sqrt{S(\frac{p}{2})}}\sin^{3}\left(\frac{p}{2}\right)e^{-\Delta\mathcal{E}\big/S(\frac{p}{2})}+o(\delta^{3})\label{eq:correction-small}\end{align}
where we define\begin{align*}
S(\tfrac{p}{2}) & =4g^{2}\frac{(r^{2}-1)^{2}}{r^{2}}+16g^{2}\sin^{2}\left(\frac{p}{2}\right)\end{align*}
and note that, in the present `small' case, $S(\tfrac{p}{2})=\frac{Q^{2}}{4\sin^{2}\left(\frac{p}{2}\right)}+16g^{2}\sin^{2}\left(\frac{p}{2}\right)\to\mathcal{E}^{2}$
when $r\to1$.

\subsubsection*{Three comments}
\begin{itemize}
\item Our result \eqref{eq:correction-small} is for the dyonic case $Q/g\sim1$.
As written it appears that $\delta\mathcal{E}\to0$ in the non-dyonic
limit $r\to1$, but this isn't correct. We've implicitly assumed,
when expanding in $\delta$, that $\delta\ll r-1\sim\sqrt{Q/g}$,
and this forbids taking $r\to1$. However, we can derive the correction
for the non-dyonic case by writing $r=1+k\delta$ before assuming
that $\delta$ is small, then expanding in $\delta$, fixing $\delta$
using the branch cut, and only then taking the limit $k\to0$. The
result is the AFZ form:\begin{align}
\delta\mathcal{E}_{r=1} & =-16g\cos(2\phi)\sin^{3}\left(\frac{p}{2}\right)e^{-2\Delta/\mathcal{E}}+o(\delta^{3}).\label{eq:AFZ-small}\end{align}

\item The condition that $\delta$ be real is that its phase $\psi$ must
be $0$ or $\pi$: \[
\psi=n\pi-\frac{p}{4}-\phi-\frac{\Delta\: Q\cot(\frac{p}{2})}{4S(\frac{p}{2})}-\frac{1}{2}\arctan\left(\frac{2\mathcal{E}}{Q}\tan(\tfrac{p}{2})\right)=0\mbox{ or }\pi.\]
 Like the energy correction \eqref{eq:correction-small}, this expression
is for the dyonic case; the phase of $\delta$ in the non-dyonic limit
$r\to1$ is instead\[
\psi_{r=1}=2\pi n-\frac{p}{2}=0\mbox{ or }\pi\]
where we have assumed $\cos(2\phi)=\pm1$. This implies $p=0\mbox{ mod }2\pi$,
which is exactly the usual condition for a closed string. 
\item Finally, notice that the the same factor $\cos(2\phi)$ appears in
these results as in the sigma-model results \eqref{eq:finite-J-RP3},
which we interpreted there as a geometric angle between adjacent magnons.
Here we can observe that for the identity \eqref{eq:Gx+G1/x-cancellation}
to hold at the evaluation point $x=X^{+}$ (in the limit $\delta\to0$,
as well as $r\to1$) , we must set $\cos(2\phi)=\pm1$. 
\end{itemize}

\subsection{Finite-size pair of small magnons}

The non-dyonic $r=1$ case for the pair of small magnons was studied
in \cite{Lukowski:2008eq}, who obtained \begin{align*}
\delta\mathcal{E}_{r=1} & =-32g\cos(2\phi)\sin^{3}\left(\frac{p}{4}\right)e^{-2\Delta/\mathcal{E}}+\ldots.\end{align*}
This result can also be obtained by adding together all the charges
of two small magnons, giving twice the correction \eqref{eq:AFZ-small}.
However the dyonic case cannot be be obtained by adding together two
dyonic finite-$J$ small magnons: they interact with each other. For
this case we must perform a similar analysis to that for the small
giant magnon above. 

The curve is $G_{u}=G_{v}=G_{\mathrm{finite}}$ and $G_{r}=0$, and
we now set \[
X^{\pm}=re^{\pm ip_{0}/4}\]
giving $p=p_{0}+\ldots$ and $\mathcal{E}=8g\frac{r^{2}+1}{2r}\sin\left(\frac{p}{4}\right)+\ldots$,
$Q=16g\frac{r^{2}-1}{2r}\sin\left(\frac{p}{4}\right)+\ldots$ .%
\footnote{As before, we give these expansions in $\delta$ in the appendix.%
} The energy correction in terms of $\delta$ reads \[
\delta\mathcal{E}=-\delta^{2}\frac{g}{2}\cos(2\phi)\frac{2r}{r^{2}+1}\sin\left(\frac{p}{4}\right)+\ldots.\]

We then fix $\delta$ using the branch cut condition connecting sheets%
\footnote{In \cite{Lukowski:2008eq} a condition for the $G_{v}$ component
to connect sheets $q_{4}$ and $q_{5}$ is used instead. (This involves
separating the two cuts slightly, so that $q_{5}\neq0$. The $G_{u}$
component has instead a cut connecting $q_{4}$ and $q_{6}$, which
gives the same equation.) The resulting condition is the same as that
given here except $n$ is replaced by $2n$. %
} $q_{4}$ and $q_{7}=-q_{4}$\begin{align*}
2\pi n & =q_{4}(x^{+})-q_{7}(x^{-})\\
 & =\dfrac{2\alpha x}{x^{2}-1}+2G_{\mathrm{finite}}^{+}(X^{+})+2G_{\mathrm{finite}}^{-}(X^{+})\end{align*}
and find the final energy correction \begin{equation}
\delta\mathcal{E}=-256g^{2}\cos(2\phi)\frac{1}{\mathcal{E}}\sin^{4}\left(\frac{p}{4}\right)e^{-\Delta\mathcal{E}\big/2S(\frac{p}{4})}+\ldots.\label{eq:correction-pair}\end{equation}
This has the same form as the $S^{5}$ string result, and exactly
matches the $RP^{3}$ magnon's correction \eqref{eq:finite-J-RP3}. 

In this case there is no difficulty about the $r\to1$ limit, where
it reduces to the $RP^{2}$ correction \eqref{eq:finite-J-RP2}. (Note
that $S(\frac{p}{4})\to\frac{1}{4}\mathcal{E}^{2}$ in this limit,
rather than $\mathcal{E}^{2}$ as in the small case.)

The phase of $\delta$ is, in this case, \[
\psi=\frac{n\pi}{2}-\frac{p}{4}-\phi-\frac{\Delta\: Q\cot(\frac{p}{4})}{8S(\frac{p}{4})}=0\mbox{ or }\pi.\]
When $Q=0$ and $\phi=0$, this means $p/2=p'=n'\pi$, $n'\in\mathbb{Z}$,
exactly matching the condition for the $RP^{2}$ magnon to be a closed
string.

\subsection{Finite-size big magnon}

The curve here is $G_{u}=G_{v}=G_{r}=G_{\mathrm{finite}}$. We write
$X^{\pm}=r\, e^{ip_{0}/4}$, and consider the `dyonic' case in the
sense that $r>1$, even though $Q=0$. Define $Q_{u}$ to be the $Q$
from the small magnon, which thanks to this choice of $p_{0}$ now
reads $Q_{u}=8g\frac{r^{2}-1}{2r}\sin\left(\frac{p}{4}\right)+\ldots$,
and we have $\mathcal{E}=8g\frac{1+r^{2}}{2r}\sin\left(\frac{p}{4}\right)+\ldots$.
Calculating the expansions of the asymptotic charges in $\delta$,
we get (as for the pair case) \[
\delta\mathcal{E}=-\delta^{2}\frac{g}{2}\cos(2\phi)\frac{2r}{r^{2}+1}\sin\left(\frac{p}{4}\right)+\ldots.\]

Now for the branch cut condition. We connect sheets $q_{3}$ and $q_{7}=-q_{4}$,
at points $x^{\pm}$ either side of the cut from $X^{+}$to $Y^{+}$,
but on the same side of the cut from $1/X^{+}$ to $1/Y^{+}$, obtaining
the matching condition \begin{align*}
2\pi n & =q_{3}(x^{+})-q_{7}(x^{-})\\
 & =2\frac{\alpha x}{x^{2}-1}+G_{\mathrm{finite}}^{+}(x)+G_{\mathrm{finite}}^{-}(x)+2G_{\mathrm{finite}}(0)-2G_{\mathrm{finite}}\left(\frac{1}{x}\right)\end{align*}
This equation fixes $\delta$, and after demanding that it be real,
we obtain the correction: \begin{align}
\delta\mathcal{E} & =-64g\cos(2\phi)\frac{r^{2}+\frac{1}{r^{2}}-2\cos(p/2)}{(r^{2}+1)(r^{2}-1)^{2}}r^{3}\sin^{3}\left(\frac{p}{4}\right)e^{-\Delta\mathcal{E}\big/S(\frac{p}{4})}+\ldots\nonumber \\
 & =-1024g^{2}\cos(2\phi)\frac{S(\frac{p}{4})}{\mathcal{E}Q_{u}^{2}}\sin^{6}\left(\frac{p}{4}\right)e^{-\Delta\mathcal{E}\big/S(\frac{p}{4})}+\ldots.\label{eq:correction-big}\end{align}

Like our small magnon result, this expression is valid only in the
dyonic case. The non-dyonic limit $r\to1$ can be approached in the
same way as for that case, by setting $r=1+k\delta$ before expanding
in $\delta$. The limit $k\to0$ then gives the result\begin{align*}
\delta\mathcal{E}_{r=1} & =-32g\cos(2\phi)\sin^{3}\left(\frac{p}{4}\right)e^{-2\Delta/\mathcal{E}}+\ldots\end{align*}
matching the $r=1$ limit of the pair of small magnons, \eqref{eq:correction-pair}
above, and thus the $RP^{2}$ string result \eqref{eq:finite-J-RP2}. 

The phase of $\delta$ in this case is \[
\psi=n\pi-\frac{p}{2}-\phi-\frac{\Delta\: Q_{u}\cot(\frac{p}{4})}{2S(\frac{p}{4})}+\arctan\left(\frac{\mathcal{E}}{Q_{u}}\tan(\tfrac{p}{4})\right).\]
As for the small case, this expression is not valid in the non-dyonic
case, where instead we get (in the case $\cos(2\phi)=\pm1$)\[
\psi_{r=1}=\frac{n\pi}{2}-\frac{p}{4}=0\mbox{ or }\pi.\]
i.e. $p=0\mbox{ mod }2\pi$, thus $p'=0\mbox{ mod }\pi$, which is
the condition for a closed string in $RP^{2}$, and matches the `pair'
case above.

\section{Conclusions}

Let us summarise the dictionary of string and curve solutions which
we have found:
\begin{itemize}
\item The small giant magnon in the curve matches the $CP^{1}$ giant magnon,
and its dyonic generalisation in $CP^{2}$.
\item The $RP^{2}$ magnon is to be identified with the pair of small magnons.
In the dyonic case this becomes the $RP^{3}$ magnon solution. 
\item The dressed solution is identified with the big giant magnon. Both
are two-parameter one-charge solutions, and when the additional parameter
($Q_{f}$ or $Q_{u}$) is sent to zero, they become the $RP^{2}$
solution / pair of small magnons, respectively. 
\end{itemize}
Note that the non-dyonic $RP^{2}$ and $CP^{1}$ string solutions
seem to have multiple descriptions in the algebraic curves: the big
and pair of small magnons differ in their excitation numbers $M_{u},M_{v},M_{r}$,
as do the two kinds of small magnons. However these numbers are all
of order $1\ll4g=\sqrt{2\lambda}$ and thus, like $Q$, invisible
in the sigma-model. In the limit $Q\to0$ the curves forget these
distinctions too.

Finite-size corrections to these magnons can be summarised as follows:
\begin{itemize}
\item In the non-dyonic cases, the corrections are always of the AFZ form.
These can be calculated in both the string and curve pictures.
\item For the $RP^{3}$ / `pair' magnon, the corrections are the same as
those for $S^{5}$ dyonic giant magnons, and can again be calculated
in both pictures.
\item For the dressed / big magnon, and also for the $CP^{2}$ / small magnon,
we have calculated corrections in the algebraic curve. These do not
have the same form as in $S^{5}$. 
\end{itemize}
Our result for the finite-$J$ corrections to the small giant magnon
differs from that of the algebraic curve calculation in \cite{Lukowski:2008eq}.
This difference can be understood to arise due to an order-of-limits
problem. As noted in section \ref{sec:Curve-Finite-J}, we need to
be carefull in the non-dyonic case with how we take limits $Q\to0$
and $\delta\to0$. However, the result of \cite{Lukowski:2008eq}
is confirmed by the Lüscher-calculations in \cite{Lukowski:2008eq,Bombardelli:2008qd}.
It would be instructive to see if these results can be explained in
a similar manner.

While the overall picture is now clear, there are various details
which it would be nice to see explicitly in the string sigma-model.
First, our $CP^{2}$ solution should certainly exist at $p\neq\pi$,
but so far we have not been able to find such solutions. Second, it
would be interesting to understand exactly how the two different $CP^{2}$
solutions join (and interact) to form one $RP^{3}$ magnon. Finally,
finite-$J$ versions of both this $CP^{2}$ solution and the dressed
solution should exist, and would provide confirmation of the energy
corrections calculated here. 

We conclude by noting that our results fit well into the context of
the integrable alternating spin-chain for operators in the ABJM gauge
theory \cite{Gaiotto:2008cg,Minahan:2008hf,Bak:2008cp,Zwiebel:2009vb,Minahan:2009te,Gromov:2008qe}.
The two small giant magnons correspond to simple magnons in either
the fundamental or anti-fundamental part of the spin-chain. The big
magnon, on the other hand, carries the same charges as the heavy scalar
excitation first discussed in \cite{Gaiotto:2008cg}. In a recent
paper Zarembo \cite{Zarembo:2009au} showed that in the BMN limit
these heavy modes disappear from the spectrum as soon as quantum corrections
are taken into account. It would be very interesting to understand
to what extent these arguments carries over to the giant magnon regime.

\section*{Acknowledgements}

We would like to thank Antal Jevicki and Joe Minahan for conversations,
and Timothy Hollowood for correspondence. 

OOS thanks the CTP at MIT for its kind hospitality during parts of
this work. MCA and IA have been supported in part by DOE grant DE-FG02-91ER40688-Task~A.
MCA thanks the Mathematics Department for financial support. IA was
also supported in part by POCI 2010 and FSE, Portugal, through the
fellowship SFRH/BD/14351/2003. The research of OOS was supported in
part by the STINT CTP-Uppsala exchange program.

\vspace{-4mm}\smash{\raisebox{-190mm}{\makebox[0cm][c]{\hspace{26cm}\includegraphics[height=8cm]{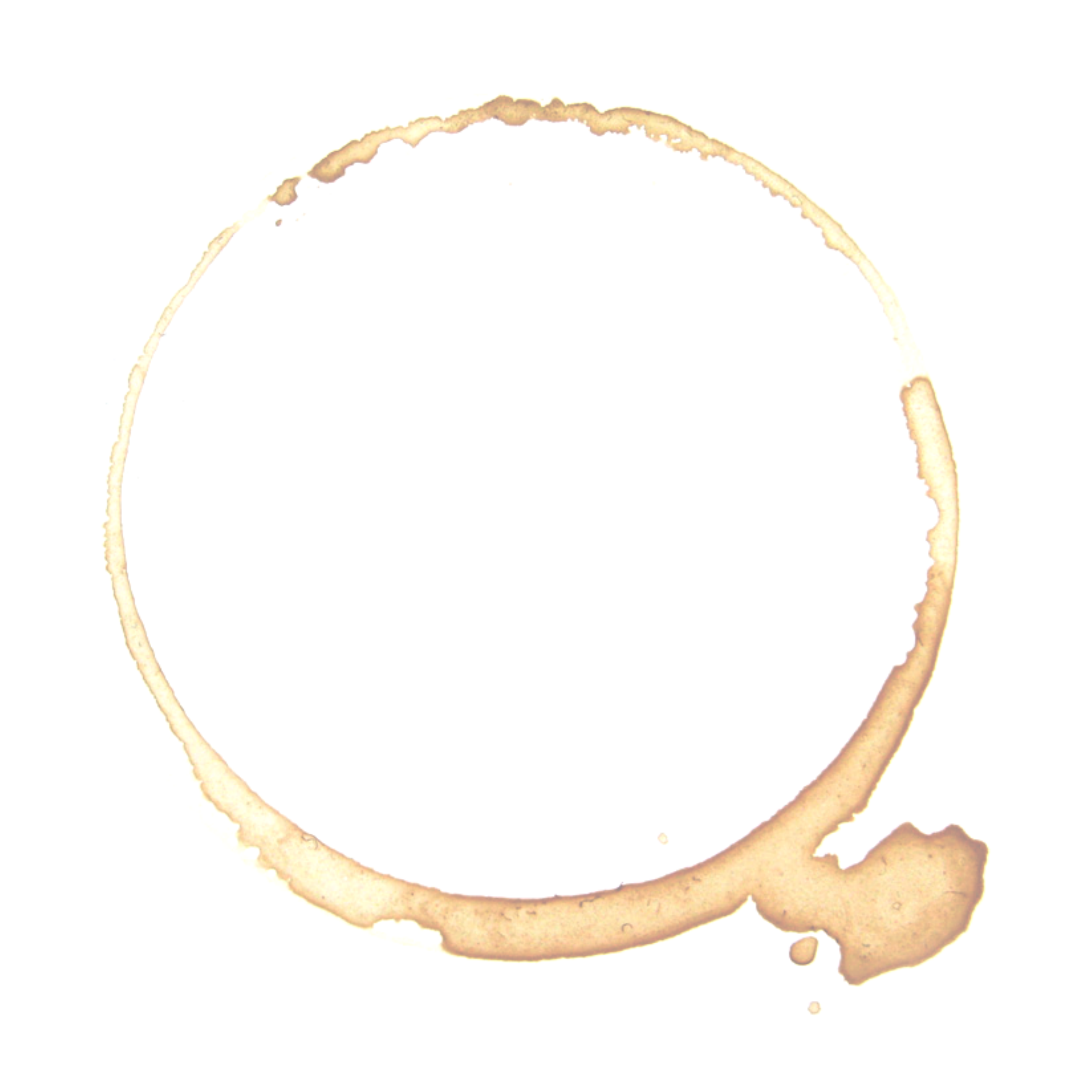}}}} 

\appendix

\section{Expansions of charges}

When working out finite-$J$ corrections to the various algebraic
curve magnons, we expanded the asymptotic charges in $\delta$, defined
by \eqref{eq:denf-Ypm} $Y^{\pm}=X^{\pm}(1\pm i\delta e^{\pm i\phi})$.
We used these expansions to work out the correction $\delta\mathcal{E}$,
for example \eqref{eq:small-deltaE-in-delta}. Here we give the expansions
of these charges explicitly.

We write all three cases at once, by setting $m=1$ for the `small'
magon and $m=2$ for the `pair' and `big'. Thus we always have $X^{\pm}=r\, e^{\pm ip/2m}$.

First, the momentum is $p=p_{0}+\delta p_{(1)}+\delta^{2}p_{(2)}+o(\delta^{3})$,
where\begin{align*}
p_{(1)} & =m\cos(\phi)\\
p_{(2)} & =\frac{3m}{8}\sin(2\phi).\end{align*}
Next, the angular momentum is $J=J_{(0)}+\delta\, J_{(1)}+\delta^{2}J_{(2)}+o(\delta^{3})$,
with\begin{align*}
J_{(0)} & =2\Delta-4gm\frac{r^{2}+1}{r}\sin\left(\frac{p_{0}}{2m}\right)\\
J_{(1)} & =-\frac{2gm}{r}\left[r^{2}\cos\left(\frac{p_{0}}{2m}+\phi\right)+\cos\left(\frac{p_{0}}{2m}-\phi\right)\right]\\
J_{(2)} & =\frac{3gm}{2r}\sin\left(\frac{p_{0}}{2m}-2\phi\right).\end{align*}
Finally, the second angular momentum is $Q=Q_{(0)}+\delta\, Q_{(1)}+\delta^{2}Q_{(2)}+o(\delta^{3})$,
where for the small and pair cases we have\begin{align*}
Q_{(0)} & =4gm\frac{r^{2}-1}{r}\sin\left(\frac{p_{0}}{2m}\right)\\
Q_{(1)} & =\frac{2gm}{r}\left[r^{2}\cos\left(\frac{p_{0}}{2m}+\phi\right)-\cos\left(\frac{p_{0}}{2m}-\phi\right)\right]\\
Q_{(2)} & =\frac{3gm}{2r}\sin\left(\frac{p_{0}}{2m}-2\phi\right).\end{align*}
For the big magnon, $Q=0$ to this order in $\delta$. We used in
the dispersion relation instead $Q_{u}$ which (as a function of $X^{\pm}$)
is the $Q$ from the small magnon. For the purpose of these expansions
(functions of $r$ and $p$) it is easier to think of this as $Q_{u}=\frac{1}{2}Q_{\mathrm{pair}}$
since the big and pair cases both have $m=2$.

\begin{small}\bibliographystyle{my-JHEP-3}
\addcontentsline{toc}{section}{\refname}\bibliography{complete-library-processed}

\providecommand{\href}[2]{#2}\begingroup\raggedright\begin{thebibliography}{10}

\bibitem{Aharony:2008ug}
O.~Aharony, O.~Bergman, D.~L. Jafferis and J.~M. Maldacena, {\it
  $\mathcal{N}=6$ superconformal {Chern--Simons}-matter theories, {M2}-branes
  and their gravity duals},  JHEP {\bf 10} (2008) 091
  [\href{http://arXiv.org/abs/0806.1218}{{arXiv:0806.1218}}].

\bibitem{Maldacena:1997re}
J.~M. Maldacena, {\it The large n limit of superconformal field theories and
  supergravity},  Adv. Theor. Math. Phys. {\bf 2} (1998) 231--252
  [\href{http://arXiv.org/abs/hep-th/9711200}{{arXiv:hep-th/9711200}}].

\bibitem{Hofman:2006xt}
D.~M. Hofman and J.~M. Maldacena, {\it Giant magnons},  J. Phys. {\bf A39}
  (2006) 13095--13118
  [\href{http://arXiv.org/abs/hep-th/0604135}{{arXiv:hep-th/0604135}}].

\bibitem{Gaiotto:2008cg}
D.~Gaiotto, S.~Giombi and X.~Yin, {\it Spin chains in $\mathcal{N}=6$
  superconformal {Chern--Simons}-matter theory},
  \href{http://arXiv.org/abs/0806.4589}{{arXiv:0806.4589}}.

\bibitem{Grignani:2008is}
G.~Grignani, T.~Harmark and M.~Orselli, {\it The {$SU(2)$} \ensuremath{\times}
  {$SU(2)$} sector in the string dual of $\mathcal{N}=6$ superconformal
  {Chern--Simons} theory},  Nucl. Phys. {\bf B810} (2008) 115--134
  [\href{http://arXiv.org/abs/0806.4959}{{arXiv:0806.4959}}].

\bibitem{Dorey:2006dq}
N.~Dorey, {\it Magnon bound states and the {AdS}/{CFT} correspondence},  J.
  Phys. {\bf A39} (2006) 13119--13128
  [\href{http://arXiv.org/abs/hep-th/0604175}{{arXiv:hep-th/0604175}}].

\bibitem{Chen:2006gea}
H.-Y. Chen, N.~Dorey and K.~Okamura, {\it Dyonic giant magnons},  JHEP {\bf 09}
  (2006) 024
  [\href{http://arXiv.org/abs/hep-th/0605155}{{arXiv:hep-th/0605155}}].

\bibitem{Ahn:2008hj}
C.~Ahn, P.~Bozhilov and R.~C. Rashkov, {\it {Neumann--Rosochatius} integrable
  system for strings on {$ {AdS}_4\times CP^3$}},  JHEP {\bf 09} (2008) 017
  [\href{http://arXiv.org/abs/0807.3134}{{arXiv:0807.3134}}].

\bibitem{Gromov:2008bz}
N.~Gromov and P.~Vieira, {\it The {$ {AdS}_4$}/{${CFT}_3$} algebraic curve},
  JHEP {\bf 02} (2008) 040
  [\href{http://arXiv.org/abs/0807.0437}{{arXiv:0807.0437}}].

\bibitem{Minahan:2006bd}
J.~A. Minahan, A.~Tirziu and A.~A. Tseytlin, {\it Infinite spin limit of
  semiclassical string states},  JHEP {\bf 08} (2006) 049
  [\href{http://arXiv.org/abs/hep-th/0606145}{{arXiv:hep-th/0606145}}].

\bibitem{Shenderovich:2008bs}
I.~Shenderovich, {\it Giant magnons in {$ {AdS}_4$}/{${CFT}_3$}: dispersion,
  quantization and finite--size corrections},
  \href{http://arXiv.org/abs/0807.2861}{{arXiv:0807.2861}}.

\bibitem{Lukowski:2008eq}
T.~{\L}ukowski and O.~Ohlsson~Sax, {\it Finite size giant magnons in the
  {$SU(2)$} \ensuremath{\times} {$SU(2)$} sector of {$ {AdS}_4\times CP^3$}},
  JHEP {\bf 12} (2008) 073
  [\href{http://arXiv.org/abs/0810.1246}{{arXiv:0810.1246}}].

\bibitem{Hollowood:2009tw}
T.~J. Hollowood and J.~L. Miramontes, {\it Magnons, their solitonic avatars and
  the {Pohlmeyer} reduction},
  \href{http://arXiv.org/abs/0902.2405}{{arXiv:0902.2405}}.

\bibitem{Kalousios:2009mp}
C.~Kalousios, M.~Spradlin and A.~Volovich, {\it On dyonic giant magnons on
  {$CP^3$}},  \href{http://arXiv.org/abs/0902.3179}{{arXiv:0902.3179}}.

\bibitem{Suzuki:2009sc}
R.~Suzuki, {\it Giant magnons on {$CP^3$} by dressing method},
  \href{http://arXiv.org/abs/0902.3368}{{arXiv:0902.3368}}.

\bibitem{Arutyunov:2006gs}
G.~Arutyunov, S.~Frolov and M.~Zamaklar, {\it Finite-size effects from giant
  magnons},  Nucl. Phys. {\bf B778} (2007) 1--35
  [\href{http://arXiv.org/abs/hep-th/0606126}{{arXiv:hep-th/0606126}}].

\bibitem{Okamura:2006zv}
K.~Okamura and R.~Suzuki, {\it A perspective on classical strings from complex
  sine-gordon solitons},  Phys. Rev. {\bf D75} (2007) 046001
  [\href{http://arXiv.org/abs/hep-th/0609026v4}{{arXiv:hep-th/0609026v4}}].

\bibitem{Astolfi:2007uz}
D.~Astolfi, V.~Forini, G.~Grignani and G.~W. Semenoff, {\it Gauge invariant
  finite size spectrum of the giant magnon},  Phys. Lett. {\bf B651} (2007)
  329--335
  [\href{http://arXiv.org/abs/hep-th/0702043}{{arXiv:hep-th/0702043}}].

\bibitem{Hatsuda:2008gd}
Y.~Hatsuda and R.~Suzuki, {\it Finite-size effects for dyonic giant magnons},
  Nucl. Phys. {\bf B800} (2008) 349--383
  [\href{http://arXiv.org/abs/0801.0747v5}{{arXiv:0801.0747v5}}].

\bibitem{Minahan:2008re}
J.~A. Minahan and O.~Ohlsson~Sax, {\it Finite size effects for giant magnons on
  physical strings},  Nucl. Phys. {\bf B801} (2008) 97--117
  [\href{http://arXiv.org/abs/0801.2064}{{arXiv:0801.2064}}].

\bibitem{Gromov:2008ec}
N.~Gromov, S.~{Sch\"{a}fer-Nameki} and P.~Vieira, {\it Efficient precision
  quantization in {AdS}/{CFT}},  JHEP {\bf 12} (2008) 013
  [\href{http://arXiv.org/abs/0807.4752}{{arXiv:0807.4752}}].

\bibitem{Sax:2008in}
O.~Ohlsson~Sax, {\it Finite size giant magnons and interactions},  Acta Phys.
  Polon. {\bf B39} (2008) 3143--3152
  [\href{http://arXiv.org/abs/0810.5236}{{arXiv:0810.5236}}].

\bibitem{Janik:2007wt}
R.~A. Janik and T.~{\L}ukowski, {\it Wrapping interactions at strong coupling:
  the giant magnon},  Phys. Rev. {\bf D76} (2007) 126008
  [\href{http://arXiv.org/abs/0708.2208}{{arXiv:0708.2208}}].

\bibitem{Bombardelli:2008qd}
D.~Bombardelli and D.~Fioravanti, {\it Finite-size corrections of the {$CP^3$}
  giant magnons: the {L\"{u}scher} terms},
  \href{http://arXiv.org/abs/0810.0704}{{arXiv:0810.0704}}.

\bibitem{Abbott:2008qd}
M.~C. Abbott and I.~V. Aniceto, {\it Giant magnons in {$ {AdS}_4\times CP^3$}:
  Embeddings, charges and a {Hamiltonian}},  JHEP {\bf 04} (2009) 136
  [\href{http://arXiv.org/abs/0811.2423}{{arXiv:0811.2423}}].

\bibitem{Grignani:2008te}
G.~Grignani, T.~Harmark, M.~Orselli and G.~W. Semenoff, {\it Finite size giant
  magnons in the string dual of $\mathcal{N}=6$ superconformal {Chern--Simons}
  theory},  JHEP {\bf 12} (2008) 008
  [\href{http://arXiv.org/abs/0807.0205}{{arXiv:0807.0205}}].

\bibitem{Lee:2008ui}
B.-H. Lee, K.~L. Panigrahi and C.~Park, {\it Spiky strings on {$ {AdS}_4\times
  CP^3$}},  JHEP {\bf 11} (2008) 066
  [\href{http://arXiv.org/abs/0807.2559v3}{{arXiv:0807.2559v3}}].

\bibitem{Ahn:2008wd}
C.~Ahn and P.~Bozhilov, {\it Finite-size effect of the dyonic giant magnons in
  $\mathcal{N}=6$ super {Chern--Simons} theory},
  \href{http://arXiv.org/abs/0810.2079}{{arXiv:0810.2079}}.

\bibitem{Harnad:1983we}
J.~P. Harnad, Y.~S. Aubin and S.~Shnider, {\it {B\"{a}cklund} transformations
  for nonlinear sigma models with values in {Riemannian} symmetric spaces},
  Commun. Math. Phys. {\bf 92} (1984) 329.

\bibitem{Sasaki:1984tp}
R.~Sasaki, {\it Explicit forms of {B\"{a}cklund} transformations for the
  {Grassmannian} and {$CP^{N-1}$} sigma models},  J. Math. Phys. {\bf 26}
  (1985) 1786.

\bibitem{Spradlin:2006wk}
M.~Spradlin and A.~Volovich, {\it Dressing the giant magnon},  JHEP {\bf 10}
  (2006) 012
  [\href{http://arXiv.org/abs/hep-th/0607009v3}{{arXiv:hep-th/0607009v3}}].

\bibitem{Gubser:2002tv}
S.~S. Gubser, I.~R. Klebanov and A.~M. Polyakov, {\it A semi-classical limit of
  the gauge/string correspondence},  Nucl. Phys. {\bf B636} (2002) 99--114
  [\href{http://arXiv.org/abs/hep-th/0204051}{{arXiv:hep-th/0204051}}].

\bibitem{Frolov:2003qc}
S.~Frolov and A.~A. Tseytlin, {\it Multi-spin string solutions in {$
  {AdS}_5\times S^5$}},  Nucl. Phys. {\bf B668} (2003) 77--110
  [\href{http://arXiv.org/abs/hep-th/0304255}{{arXiv:hep-th/0304255}}].

\bibitem{Kazakov:2004qf}
V.~A. Kazakov, A.~Marshakov, J.~A. Minahan and K.~Zarembo, {\it Classical /
  quantum integrability in {AdS}/{CFT}},  JHEP {\bf 05} (2004) 024
  [\href{http://arXiv.org/abs/hep-th/0402207}{{arXiv:hep-th/0402207}}].

\bibitem{Kazakov:2004nh}
V.~A. Kazakov and K.~Zarembo, {\it Classical / quantum integrability in
  non-compact sector of {AdS}/{CFT}},  JHEP {\bf 10} (2004) 060
  [\href{http://arXiv.org/abs/hep-th/0410105}{{arXiv:hep-th/0410105}}].

\bibitem{Beisert:2004ag}
N.~Beisert, V.~A. Kazakov and K.~Sakai, {\it Algebraic curve for the so(6)
  sector of {AdS}/{CFT}},  Commun. Math. Phys. {\bf 263} (2006) 611--657
  [\href{http://arXiv.org/abs/hep-th/0410253}{{arXiv:hep-th/0410253}}].

\bibitem{SchaferNameki:2004ik}
S.~{Sch\"{a}fer-Nameki}, {\it The algebraic curve of 1-loop planar
  $\mathcal{N}=4$ {SYM}},  Nucl. Phys. {\bf B714} (2005) 3--29
  [\href{http://arXiv.org/abs/hep-th/0412254}{{arXiv:hep-th/0412254}}].

\bibitem{Beisert:2005bm}
N.~Beisert, V.~A. Kazakov, K.~Sakai and K.~Zarembo, {\it The algebraic curve of
  classical superstrings on {$ {AdS}_5\times S^5$}},  Commun. Math. Phys. {\bf
  263} (2006) 659--710
  [\href{http://arXiv.org/abs/hep-th/0502226}{{arXiv:hep-th/0502226}}].

\bibitem{Beisert:2005di}
N.~Beisert, V.~A. Kazakov, K.~Sakai and K.~Zarembo, {\it Complete spectrum of
  long operators in $\mathcal{N}=4$ {SYM} at one loop},  JHEP {\bf 07} (2005)
  030 [\href{http://arXiv.org/abs/hep-th/0503200}{{arXiv:hep-th/0503200}}].

\bibitem{Stefanski:2008ik}
B.~Stefanski, {\it {Green--Schwarz} action for type {IIA} strings on {$
  {AdS}_4\times CP^3$}},  Nucl. Phys. {\bf B808} (2009) 80--87
  [\href{http://arXiv.org/abs/0806.4948}{{arXiv:0806.4948}}].

\bibitem{Arutyunov:2008if}
G.~Arutyunov and S.~Frolov, {\it Superstrings on {$ {AdS}_4\times CP^3$} as a
  coset sigma-model},  JHEP {\bf 09} (2008) 129
  [\href{http://arXiv.org/abs/0806.4940}{{arXiv:0806.4940}}].

\bibitem{Vicedo:2007rp}
B.~Vicedo, {\it Giant magnons and singular curves},  JHEP {\bf 12} (2007) 078
  [\href{http://arXiv.org/abs/hep-th/0703180}{{arXiv:hep-th/0703180}}].

\bibitem{Minahan:2008hf}
J.~A. Minahan and K.~Zarembo, {\it The {Bethe} ansatz for superconformal
  {Chern--Simons}},  JHEP {\bf 09} (2008) 040
  [\href{http://arXiv.org/abs/0806.3951}{{arXiv:0806.3951}}].

\bibitem{Bak:2008cp}
D.~Bak and S.-J. Rey, {\it Integrable spin chain in superconformal
  {Chern--Simons} theory},  JHEP {\bf 10} (2008) 053
  [\href{http://arXiv.org/abs/0807.2063}{{arXiv:0807.2063}}].

\bibitem{Zwiebel:2009vb}
B.~I. Zwiebel, {\it Two-loop integrability of planar $\mathcal{N}=6$
  superconformal chern- simons theory},
  \href{http://arXiv.org/abs/0901.0411}{{arXiv:0901.0411}}.

\bibitem{Minahan:2009te}
J.~A. Minahan, W.~Schulgin and K.~Zarembo, {\it Two loop integrability for
  {Chern--Simons} theories with $\mathcal{N}=6$ supersymmetry},
  \href{http://arXiv.org/abs/0901.1142}{{arXiv:0901.1142}}.

\bibitem{Gromov:2008qe}
N.~Gromov and P.~Vieira, {\it The all loop {$ {AdS}_4$}/{${CFT}_3$} {Bethe}
  ansatz},  \href{http://arXiv.org/abs/0807.0777}{{arXiv:0807.0777}}.

\bibitem{Zarembo:2009au}
K.~Zarembo, {\it Worldsheet spectrum in {$ {AdS}_4$}/{${CFT}_3$}
  correspondence},  \href{http://arXiv.org/abs/0903.1747}{{arXiv:0903.1747}}.

\end{thebibliography}\endgroup

\end{small}
\end{document}